\documentclass[smallcondensed]{svjour3}
\smartqed
\usepackage{cite}
\usepackage{amsmath,amssymb,amsfonts,dsfont}
\usepackage{algorithmic}
\usepackage{graphicx}
\usepackage{hyperref}
\usepackage{textcomp}
\usepackage{xcolor}
\usepackage{pifont}
\usepackage{textpos}
\usepackage{multirow}
\newcommand{\xmark}{\ding{55}}
\def\BibTeX{{\rm B\kern-.05em{\sc i\kern-.025em b}\kern-.08em
    T\kern-.1667em\lower.7ex\hbox{E}\kern-.125emX}}
    
\newcommand{\liu}[1]{\textcolor{black}{#1}}
    
\begin{document}

\title{From Distributed Machine Learning to Federated Learning: A Survey}

\author{Ji Liu\footnotemark[5],
        Jizhou Huang\footnotemark[5],
        Yang Zhou\footnotemark[6],
        Xuhong Li\footnotemark[5],
        Shilei Ji\footnotemark[5],
        Haoyi Xiong\footnotemark[5],
        Dejing Dou\footnotemark[1]\footnotemark[5]\footnotemark[7]
}

\date{Received: date / Accepted: date}

\institute{
\footnotemark[1] Corresponding author. \\
\footnotemark[5] Baidu Inc., Beijing, China.\\
\footnotemark[6] Computer Science and Software Engineering Department, Auburn University, Alabama, United States.\\
\footnotemark[7] Computer and Information Science Department, University of Oregon, United States.
}

\authorrunning{Liu et al.}

\maketitle

\begin{textblock*}{8cm}(3cm,15cm) 
   {\huge To appear in KAIS}
\end{textblock*}

\begin{abstract}
In recent years, data and computing resources are typically distributed in the devices of end users, various regions or organizations. Because of laws or regulations, the distributed data and computing resources cannot be \liu{aggregated or} directly shared among different regions or organizations for machine learning tasks. Federated learning emerges as an efficient approach to exploit distributed data and computing resources, so as to collaboratively train machine learning models\liu{. At the same time, federated learning obeys} the laws and regulations and ensures data security and data privacy. In this paper, we provide a comprehensive survey of existing works for federated learning. \liu{First, we propose a functional architecture of federated learning systems and a taxonomy of related techniques. Second, we explain the federated learning systems from four aspects: diverse types of parallelism, aggregation algorithms, data communication, and the security of federated learning systems. Third, we present four widely used federated systems based on the functional architecture. Finally, we summarize the limitations and propose future research directions.}
\end{abstract}

\section{Introduction}

With billions of connected Internet of Things (IoT) devices \cite{abou2019co}, smartphones \cite{ochiai2019real} and large websites around the world, in recent years, we have witnessed huge amounts of data generated and dispersed over various mobile devices of end users, or the data centers of different organizations. As the data contain sensitive information of end users or organizations, such as facial images, location-based services, health information \cite{lyu2020threats}, or personal economic status, moving the raw data from personal devices or data centers of multiple organizations to a centralized server or data center may pose immediate or potential information leakage. Due to the concerns of data security and data privacy, legal restrictions, such as the Cybersecurity Law of the People’s Republic (CLPR) of China \cite{CCL}, the General Data Protection Regulation (GDPR) \cite{GDPR} in European Union, the Personal Data Protection Act (PDP) \cite{chik2013singapore} in Singapore, the California Consumer Privacy Act (CCPA) \cite{CCPA}, and the Consumer Privacy Bill of Rights (CPBR) \cite{Gaff2014} in the United States, have been introduced and put in practice, which makes data aggregation from distributed devices, multiple regions, or organizations, almost impossible \cite{yang2019federated}. In addition, computing and storage resources are also typically distributed in multiple regions \cite{liu2018efficient} and organizations \cite{liaqat2017federated}, which cannot be aggregated in a single data center.

Federated Learning (FL) emerges as an efficient approach to exploit the distributed resources to collaboratively train a machine learning model. FL is a distributed machine learning approach where multiple users collaboratively train a model, while keeping the raw data decentralized without being moved to a single server or data center \cite{kairouz2019advances, yang2019federated}.
FL not only exploits the distributed resources to efficiently carry out the training process of machine learning, but also promises to provide security and privacy for the decentralized raw data. Within FL, the raw data, or the data generated based on the raw data with security processing, serves as the training data. FL only allows the intermediate data to be transferred among the distributed computing resources while avoiding the transfer of training data. 
The distributed computing resources refer to mobile devices of end users or servers of multiple organizations.
FL brings the code to the data, instead of bringing the data to the code, and it addresses the fundamental problems of privacy, ownership, and locality of data \cite{mcmahan2017communication}.
In this way, FL can enable multiple users to train a model while satisfying the legal data restrictions.

Traditional centralized machine learning approaches typically gather the distributed raw data generated on different devices or organizations to a single server or a cluster with shared data storage, which may bring serious data privacy and security concerns \cite{zhu2021}. 
The centralized approaches, in general, are associated with diverse challenges, including computational power and training time, and most importantly, security and privacy with respect to distributed data \cite{mothukuri2021survey}.
FL differs from the centralized approach in three aspects. First, FL does not allow direct raw data communication, while the centralized approach has no restriction. Second, FL exploits the distributed computing resources in multiple regions or organizations, while the centralized approach generally only utilizes a single server or a cluster in a single region, which belongs to a single organization. Third, FL generally takes advantage of encryption or other defense techniques to ensure the data privacy or security, while the centralized approach pays little attention to these security issues \cite{zhu2021}.

The term ``federated learning'' was first introduced in 2016 \cite{mcmahan2017communication}, which focuses on the unbalanced and non-Independent and Identically Distributed (non-IID) data in mobile devices. 
The concept of FL was extended to three data scenarios, i.e., horizontal, vertical, and hybrid \cite{yang2019federated, zhu2021}. 
The horizontal FL addresses the decentralized data of the same features, while the identifications are different. The vertical FL handles the decentralized data of the same identifications with different features. The hybrid FL deals with the data of different identifications and different features. Then, FL is formally defined as a machine learning approach where multiple clients collaborate in solving a machine learning problem while the raw data is stored locally and is neither exchanged nor transferred \cite{kairouz2019advances}. 

An FL system is an efficient tool to carry out FL with decentralized data and resources. Several open-source FL systems, e.g., FATE \cite{FATE}, PaddleFL \cite{PaddleFL}, TensorflowFL \cite{TFF}, and Pysyft \cite{Pysyft}, are now intensively used by both research communities, e.g., healthcare \cite{brisimi2018federated}, and computer visions \cite{liu2020fedvision, hefedcv}, and by industrial groups, e.g., WeBank \cite{FLWhitePaper}. Although various FL systems exist, the architecture of FL systems has common features: In particular, they share the capability to collaboratively train a machine learning model. Most FL systems are composed of four layers, i.e., presentation, user services, FL training, and infrastructure. These four layers enable FL system users to design, execute, and analyze machine learning models with distributed data.

Although FL differs from the centralized machine learning approaches, it not only utilizes novel techniques designed for FL, but also takes advantage of the techniques designed for distributed machine learning. 
FL exploits parallelization techniques designed for distributed machine learning.
For instance, horizontal FL exploits the data parallelism, which trains multiple instances of the same model on different subsets of the training dataset \cite{verbraeken2020survey}. Vertical FL utilizes model parallelism to distribute parallel paths of a single model to multiple devices in order to handle the data of different features  \cite{verbraeken2020survey}. 
Multiple aggregation algorithms \cite{chen2019communication} are proposed to aggregate the models in distributed computing resources.
Data transfer techniques are also utilized in FL, e.g., model compression \cite{caldas2018expanding}. 
As FL promises to provide data security and data privacy, diverse defense techniques, e.g., differential privacy \cite{mcmahan2017learning}, homomorphic encryption \cite{hardy2017private}, and Robustness Aggregation \cite{pillutla2019robust}, are designed to address the possible attacks \cite{wang2019beyond, hitaj2017deep, geiping2020inverting}.

There have been a few surveys of FL. Some works \cite{yang2019federated, kairouz2019advances, li2019survey} provide a comprehensive study of FL, from the taxonomy of FL to the techniques, e.g., the efficiency, data privacy, security, and applications of FL. Some surveys \cite{mothukuri2021survey, li2019survey, lyu2020threats} focus on the data privacy and security of FL. Other surveys present the application of FL in a specific area, e.g., healthcare informatics \cite{xu2020federated}, mobile edge networks \cite{lim2020federated}, and neural architecture searches \cite{zhu2021}, and they personalize global models to work better for individual clients \cite{kulkarni2020survey}. However, few of them present the architecture of FL or analyze parallelization techniques in FL.

In this paper, we provide a survey of federated learning and the related parallelization techniques. The main contributions of this paper are:
\begin{itemize}
    \item A four-layer FL system architecture, which is useful for discussing the techniques for FL. This architecture can also be a baseline for other work and can help with the assessment and comparison of FL systems.
    \item A taxonomy of FL-related techniques, including the parallelization techniques, the aggregation algorithms, and the techniques for data communication and security, with a comparative analysis of the existing solutions.
    \item A discussion of research issues to improve the efficiency and security of FL systems.
\end{itemize}

This paper is organized as follows. Section \ref{sec:overview} gives an overview of the execution of FL, including the FL system architectures and basic functional architecture of FL systems. Section \ref{sec:pexec} focuses on the techniques used for distributed training of FL and aggregation methods. Section \ref{sec:security} presents the techniques for distributed execution, data communication and data security of FL. Section \ref{sec:frameworks} demonstrates the existing FL frameworks.
Section \ref{sec:future} discusses the open issues raised for the execution of FL with distributed resources. Section \ref{sec:con} summarizes the main findings of this study.

\section{An Overview of Federated Learning}
\label{sec:overview}

In this section, we introduce the basic concepts of federated learning. Then, we present the life cycle of FL models. Afterwards, we detail the functional architecture and the corresponding functionality of FL systems.

\subsection{Basic Concepts}

Machine learning is the process to automatically extract the models or patterns from data \cite{goodfellow2016deep}. 
The models or patterns are expressed as machine learning models.
A machine learning model is an ensemble of a model structure, which is typically expressed as a Directed Acyclic Graph (DAG), data processing units, e.g., activation functions in Deep Neural Networks (DNNs), and the associated parameters or hyper-parameters. 
The input data can be processed through a machine learning model to generate the output, e.g., the prediction results or the classification results, which is the inference process.
The machine learning model is generated based on the training data, which is the training process.
During the training process, the parameters or the model structure of the machine learning model \cite{He2020CVPR, he2020towards} is adjusted based on a training algorithm in order to improve the performance, e.g., the accuracy or the generalization capacity. The training algorithm is also denoted by machine learning algorithms. The duration of the training process is training time.

According to whether the training data have labels, the training process of machine learning can be classified into four types \cite{verbraeken2020survey}, i.e., supervised learning, unsupervised learning, semi-supervised learning, and reinforcement learning. Supervised learning represents that a machine learning task exploits the training data composed of input features and the corresponding labels \cite{verbraeken2020survey}. In this paper, we focus on this type of training data. For instance, each data point in the training dataset contains $(x,y)$, where $x$ represents the input features and $y$ represents the desired output value. Unsupervised learning represents that a machine learning task exploits the training data, which only consists of input features without output values; i.e., each data point only contains $x$ and does not have $y$. Semi-supervised learning represents that one (generally small) part of the training data contains output values, while the other (generally small) part of the training does not. Reinforcement learning represents that each iteration in the training process considers its observation of the environment from the last iteration. 

While the training data become huge, e.g., on the order of terabyte \cite{canini2012sibyl}, or when the training data is inherently distributed or too big to store on single machines \cite{verbraeken2020survey}, the training process is carried out using distributed resources, which is distributed machine learning. One of the important features of the distributed machine learning is that it can significantly accelerate the training speed so as to reduce the training time. Diverse parallelization techniques are used in distributed machine learning. For instance, Graphics Processing Units (GPUs) using Single Instruction Multiple Data (SIMD) \cite{flynn1972some} and Tensor Processing Units (TPUs) using Multiple Instructions Multiple Data (MIMD) \cite{flynn1972some} are exploited \cite{verbraeken2020survey}. In addition, distributed machine learning takes advantage of three types of parallelism to parallelize the training process, i.e., data parallelism \cite{verbraeken2020survey}, model parallelism \cite{verbraeken2020survey}, and pipeline parallelism  \cite{huang2018gpipe, He2021PipeTransformer, liu2015survey}. With the data parallelism approach, the training data is partitioned as many times as the number of computing resources, and all computing resources subsequently apply the same machine learning algorithm to process different chunks of the data sets \cite{verbraeken2020survey}. With the model parallelism approach, exact copies of the entirety of the data (the training data or the intermediate data)
are processed by each computing resource, each of which exploits different parts of the machine learning model \cite{verbraeken2020survey}. The pipeline parallelism approach combines the data parallelism and the model parallelism. With this approach, each computing resource processes a part of the training data with a part of the machine learning model, while the processing, e.g., computation or communication, at each node can be parallelized \cite{narayanan2019pipedream}. 

FL is a distributed machine learning approach where multiple users collaboratively train a model, while keeping the raw data distributed without being moved to a single server or data center \cite{kairouz2019advances, yang2019federated}.
The model used for FL is denoted by FL model.
FL is first proposed to handle the unbalanced and non-Independent and Identically Distributed (non-IID) data of the same features in mobile devices \cite{mcmahan2017communication}.
Then, the concept of FL is extended to the distributed data of diverse features in multiple organizations \cite{yang2019federated} or various regions \cite{mcmahan2021advances}.
FL systems are used within one or multiple phases of the life cycle of FL models. An FL system is a distributed system to manage the distributed training process with distributed resources.

FL is a special type of distributed machine learning, which differs from other distributed machine learning approaches in the following three points.
First, FL does not allow direct raw data communication, while other approaches have no restriction. 
As the raw data are of multiple ownerships, FL approaches with this restriction can meet the requirements defined by the related laws, e.g., CLPR \cite{CCL}, GDPR \cite{GDPR}, PDPA \cite{chik2013singapore}, CCPA \cite{CCPA}, and CPBR \cite{Gaff2014}.
In particular, the consent (GDPR Article 6) and the data minimalization principle (GDPR Article 5) limit data collection and storage to only what is consumer-consented and what is absolutely necessary for processing \cite{lim2020federated}.
Second, FL exploits the distributed computing resources in multiple regions or organizations, while the other approaches generally only utilize a single server or a cluster in a single region, which belongs to a single organization. FL enables the collaboration among multiple organizations.
Third, FL generally takes advantage of encryption or other defense techniques to ensure the data privacy or security, while the other approaches pay little attention to this security issue \cite{zhu2021}. FL promises to ensure the privacy and security of the raw data, as the leakage of information may incur significant financial \cite{UberLoss,GoogleLoss} and reputational \cite{FacebookLoss} losses. 

During the training process of FL, an optimization problem is solved as shown in Formula \ref{eq:problem}. Given $n$ training dataset $\mathcal{D} = {D_1, D_2, ..., D_n}$, where each data point $(x,y)\sim\mathcal{D}$, the problem of FL is to learn a function $\widehat F$ from all possible hypotheses $\mathcal{H}$, while minimizing the expectation of loss over the distribution of all the dataset $\mathcal{D}$. 
\begin{equation}
\label{eq:problem}
    \widehat F = \underset{F\in\mathcal{H}}{\mathrm{argmin}} \underset{(x,y)\in\mathcal{D}}{\mathbb{E}} L(y, F(x)),
\end{equation}
where $L(y, F(x))$ refers to the loss of $F(x)$ to the label $y$. 
During the training process, the Stochastic Gradient Descent (SGD) approach \cite{robbins1951stochastic, zinkevich2010parallelized} is generally used to minimize the loss function using Formula \ref{eq:gd}.
\begin{equation}
\label{eq:gd}
    F_{k+1}(x)\gets F_k(x) - \eta_k \nabla F_k(x),
\end{equation}
where $F_k(x)$ refers to the learned model in the $k^{th}$ iteration,  $\nabla F_k(x)$ refers to the gradient of the model at the $k^{th}$ iteration based on the model already obtained $F_k(x)$ and the training dataset, $\eta_k$ refers to the learning rate, and $F_{k+1}(x)$ refers to the update model of the $k^{th}$ iteration. Within each iteration, there are two phases, i.e., forward propagation and backward propagation. The forward propagation calculates the output based on the input data $x$ using the model, while the backward propagation calculates the gradients $\nabla F_k(x)$ and updates the model.
When the calculation is distributed among multiple computing resources, the gradients or models of each computing resource are aggregated using an aggregation algorithm (see details in Section \ref{sebsec:aggre}), in order to achieve consensus of multiple models and to generate a global model. \liu{The learning rate can be dynamically adapted using a local adaptive optimizer, e.g., Adam, and/or cross-round learning rate schedulers \cite{zhang2021federated}.}

\subsection{FL Model Life Cycle}

The life cycle of an FL model is a description of the state transitions of an FL model from creation to completion \cite{liu2015survey, kairouz2019advances}. Lo \textit{et al.} \cite{lo2021architectural} propose that the life cycle of an FL model consists of 8 phases: initiated, broadcast, trained, transmitted, aggregated, evaluated, deployed, and monitored. Kairouz \textit{et al.} \cite{kairouz2019advances} propose that the life cycle of an FL model includes 6 phases: problem identification, client instrumentation, simulation prototyping, federated model training, model evaluation, and deployment. However, they focus on the FL with distributed data in mobile devices. In this paper, we adopt a combination of workflow life cycle views \cite{lo2021architectural,kairouz2019advances} with a few variations \cite{yang2019federated,liu2015survey}, condensed into four phases:
\begin{enumerate}
    \item The composition phase \cite{kairouz2019advances} is for the creation of an FL model, which is used to address a specific machine learning problem, e.g., classification. First, a machine learning model is created to address the problem with certain requirements, e.g., the requirement of accuracy. Then, the machine learning model is adapted to FL scenarios. For instance, if the distributed data is of different features, the machine learning model is partitioned (see details in Section \ref{subsubsec:modelP}) to process the distributed data. 
    \item The FL training phase \cite{kairouz2019advances, lo2021architectural, yang2019federated} is for the training phase of the FL model. During this phase, a training strategy, which includes parallelism and aggregation algorithms (see details in Section \ref{sec:pexec}), is used to update the parameters, hyper parameters, and even the structure of the network, in order to improve the accuracy and the generalization capacity of the FL model.  
    \item The FL model evaluation phase \cite{kairouz2019advances, liu2015survey} is to apply the trained FL models, in order to analyze the performance of the trained FL models on a simulation platform or a real distributed system \cite{he2020fedml}. As a result, the FL models with the best performance are selected. If the FL models do not meet the requirements, the FL model should be modified, or the training phase should be carried out again.
    \item The FL model deployment phase \cite{kairouz2019advances} is to deploy the FL model in a real-life scenario to process the data. If the final model can be shared without restriction, there is no difference between the FL model deployment and the model generated from a traditional centralized approach. Otherwise, the deployment of the final model should consider the ownership of the corresponding parts. 
\end{enumerate}

\subsection{Functional Architecture of FL Systems}

\begin{figure}[htbp]
    \centering
    \includegraphics[width=0.8\textwidth]{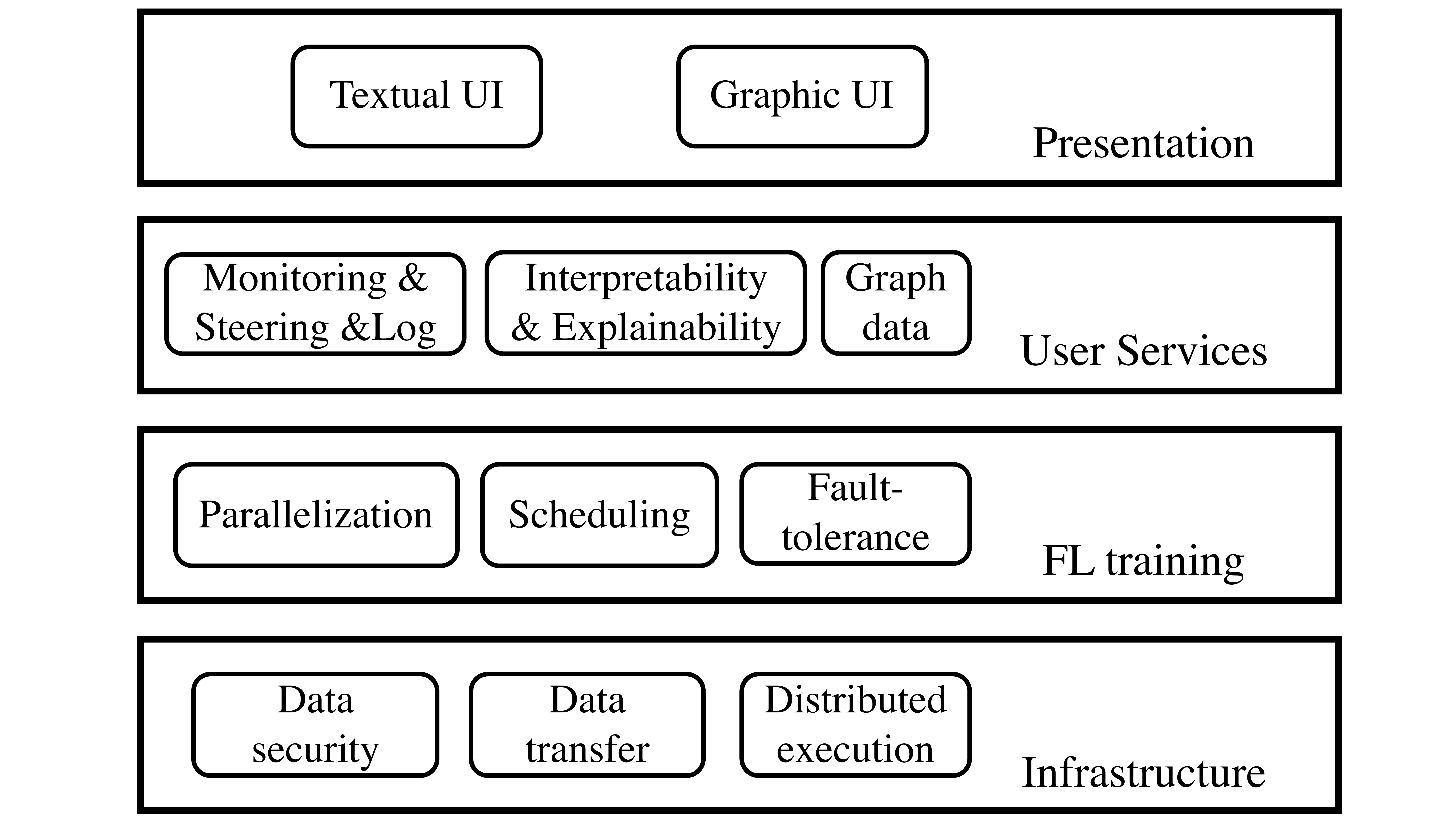}
    \caption{\liu{Functional architecture of an FL system.}}
    \label{fig:layers}
\end{figure}

The functional architecture of an FL system can be layered as follows \cite{liu2015survey}: presentation, user services, FL training, and infrastructure. Figure \ref{fig:layers} shows this architecture. The higher layers exploit the lower layers to provide their own functionality. A user interacts with an FL system through the presentation layer and realizes independent functionalities at the user services layer. During the training phase of FL models, a Federated Learning Execution Plan (FLEP) is generated, and the corresponding distributed training is carried out at the FL training layer. The FLEP is composed of a type of parallelism, a scheduling strategy, and a fault-tolerance mechanism. The FL system manages the physical resources through the infrastructure layer for the distributed training. 

\subsubsection{Presentation Layer}

The presentation layer is a User Interface (UI) for the interaction between Users and FL systems at one or multiple stages of the FL model life cycle. The UI can be textual or graphical. This interface is responsible for designing a new FL model or choosing an existing machine learning model as an FL model. In addition, this layer also supports the modules at the user services layer, e.g., shows the status of the distributed training process. The textual UI is largely used for designing FL models \liu{based on the command line or scripts \cite{Lin2021FedNLP}}. The models can be directly expressed using Python, with the textual interface in PaddleFL \cite{PaddleFL}, TensorFlowFL \cite{TFF}, PySyft \cite{ryffel2018generic}, and FATE \cite{FATE}. A graphic UI can make the interaction more practical, while the users can drag or drop the data processing element to design an FL model. For instance, FATE \cite{FATE} provides a Graphic UI (GUI) through a web portal. However, the graphic portal also exploits textual programming languages as inner representations of an FL model. 

\subsubsection{User Services Layer}

The user service layer supports the expected functionalities, i.e., \liu{monitoring and steering and log; interpretability and explainability; and graph data.} The monitoring enables the users to get the real-time status of the distributed training process. As the training process of FL models can be very long, e.g., from several hours to days \cite{kairouz2019advances}, it is of much importance to track the execution status, which allows the user to verify whether the training proceeds normally. The log service is generally supported by major FL systems, which can be used to analyze the training process. In addition, the log generated during the training process can be used to debug the system or adjust the FL model. FATE provides a visual monitoring board to users through its GUI. When there are unexpected results or errors during the training process, steering enables users to adjust the training process in order to reduce the time necessary to carry out the distributed training from scratch. Most FL systems can enable the users to stop the training, while the adjustment of parameters is not fully supported by major FL systems. The interpretability of FL is to describe the internals of an FL system in a way that is understandable to humans \cite{gilpin2018explaining}. The explainability focuses on explaining the representation of data inside an FL model \cite{gilpin2018explaining}. With interpretability and explainability, the FL system can \liu{provide} a description of the results of the trained FL model based on the training data and the distributed training process. Shapley values have been used to provide the interpretability \cite{wang2019interpret}, while both the interpretability and explainability remain open challenges as each is hard to fully support. 

\liu{In the real world, graph data widely exist in multiple domains, and a bunch of FL approaches have been proposed to handle the decentralized graph data for community detection \cite{Ke2021Federated}, financial crime \cite{Suzumura2019Towards}, and especially knowledge graph completion \cite{Chen2020FedE}.
FL is particularly useful in the field of knowledge graph completion, as a knowledge graph could not only contain text but also images or other type of data, i.e., multimodal Knowledge Graphs \cite{zhao2021multimodal}; and the completion is realized in a collaborative fashion within an FL system \cite{lin2020improving}.
The decentralized graphs can be inter-graph, i.e., the decentralized data belongs to multiple graphs, or intra-graph, i.e., the decentralized data is within one big graph \cite{Zhang2021FederatedGraph}, while most of the existing works focus on the intra-graph situation.
Horizontal FL techniques can be exploited on the Graph Neural Networks (GNN) \cite{wu2021fedgnn, Meng2021Cross} with encryption techniques \cite{jiang2020federated} (see details in Section \ref{subsubsec:infrastructure}) while the performance of FL may be much worse than that of centralized GNNs \cite{he2021fedgraphnn}.
The aggregation algorithms (see details in Section \ref{subsubsec:centralized}) are also proposed based on the FedAvg \cite{zhao2021multimodal} or optimal transportation \cite{lin2020improving}. 
In addition, decentralized aggregation algorithms (see details in Section \ref{subsubsec:decentralized}) are also proposed to deal with the decentralized graph data for social network \cite{He2019Central} and drug discovery \cite{He2021SpreadGNN}.
While the fine-tuning of the FL system is time-consuming, Bayesian optimization \cite{Zheng2021ASFGNN} and evolutionary optimization strategies \cite{Wang2021AGCNS} are utilized to automatically tune the hyper-parameters and the network structure, respectively.  
Graph data can be vertically distributed where the features of the nodes are distributed across multiple data owners, and a data owner may or may not have the edges. Vertical FL exploits embeddings \cite{zhou2020vertically,Chen2020FedE} or autoencoders \cite{zhao2021multimodal} to represent the nodes, which can be transferred to train a GNN. In addition, the differential privacy (see details in Section \ref{subsubsec:infrastructure}) is combined with the embeddings to protect the data privacy \cite{peng2021federated} of knowledge graphs. }

\subsubsection{FL Training Layer}

The FL training layer carries out the distributed training process with distributed data and computing resources. This layer consists of three modules: parallelization, scheduling, and fault-tolerance. FL parallelization exploits diverse types of parallelism, e.g., data parallelism, model parallelism, and pipeline parallelism, to generate executable tasks. Through the FL scheduling module, an FL system produces a Scheduling Plan (SP) of executable tasks, which aims at fully exploiting distributed computing resources and preventing training stalling. During the training process, the SP is generally defined by a training algorithm, which aggregates the updates, i.e., gradients or models, from each computing resource in order to generate a final machine learning model. The FL fault-tolerance mechanism handles the failures or errors of task execution and the connection of distributed resources. Reactive approaches are generally exploited, e.g., using check-points, restart, and task replication \cite{Bonawitz19}. A reactive approach reduces the effect of failures after perceiving failures \cite{ganga2013fault}. An FLEP, which captures the execution directives, typically the result of compiling and optimizing the training process of FL models, is generated at this layer.

\subsubsection{Infrastructure Layer}
\label{subsubsec:infrastructure}

The infrastructure layer provides the interaction between an FL system and the distributed resources, including the computing resources, storage resources, network resources, and data resources. This layer contains three modules: a data security module, a data transfer module, and a distributed execution module. The data security module generally exploits Differential Privacy (DP) \cite{abadi2016deep} and encryption techniques, e.g., homomorphic \cite{aono2017privacy}, to protect the raw data used during the training process. Although the raw data cannot be directly transferred, intermediate data, e.g., the gradients or models, can be communicated among distributed computing resources. The data transfer module exploits data compression techniques \cite{Stich2018} to improve the data transfer efficiency. 
At this layer, the FLEP generated at the FL training layer is carried out within the distributed execution module; i.e., concrete tasks are executed in distributed computing resources.

\section{Distributed Training}
\label{sec:pexec}

In this section, we present the distributed training process for FL. First, we present three types of parallelism in distributed training and the application within FL. The parallelism approaches are generally implemented in the parallelization module. Then, we discuss existing aggregation algorithms for the distributed training, which is implemented in the scheduling module.

\subsection{Parallelism \& FL Types}

Three types of parallelism exist for distributed machine learning: data parallelism, model parallelism, and pipeline parallelism \cite{verbraeken2020survey, liu2015survey}. FL can be classified to three types, i.e., horizontal, vertical, and hybrid \cite{zhu2021,yang2019federated}. The horizontal FL generally exploits data parallelism, and the vertical FL typically takes advantage of model parallelism. However, the hybrid FL relies on transfer learning \cite{pan2009survey}, which is not a parallelism approach and is out of the scope of this paper. 

\begin{figure}[htbp]
    \centering
    \includegraphics[width=0.65\textwidth]{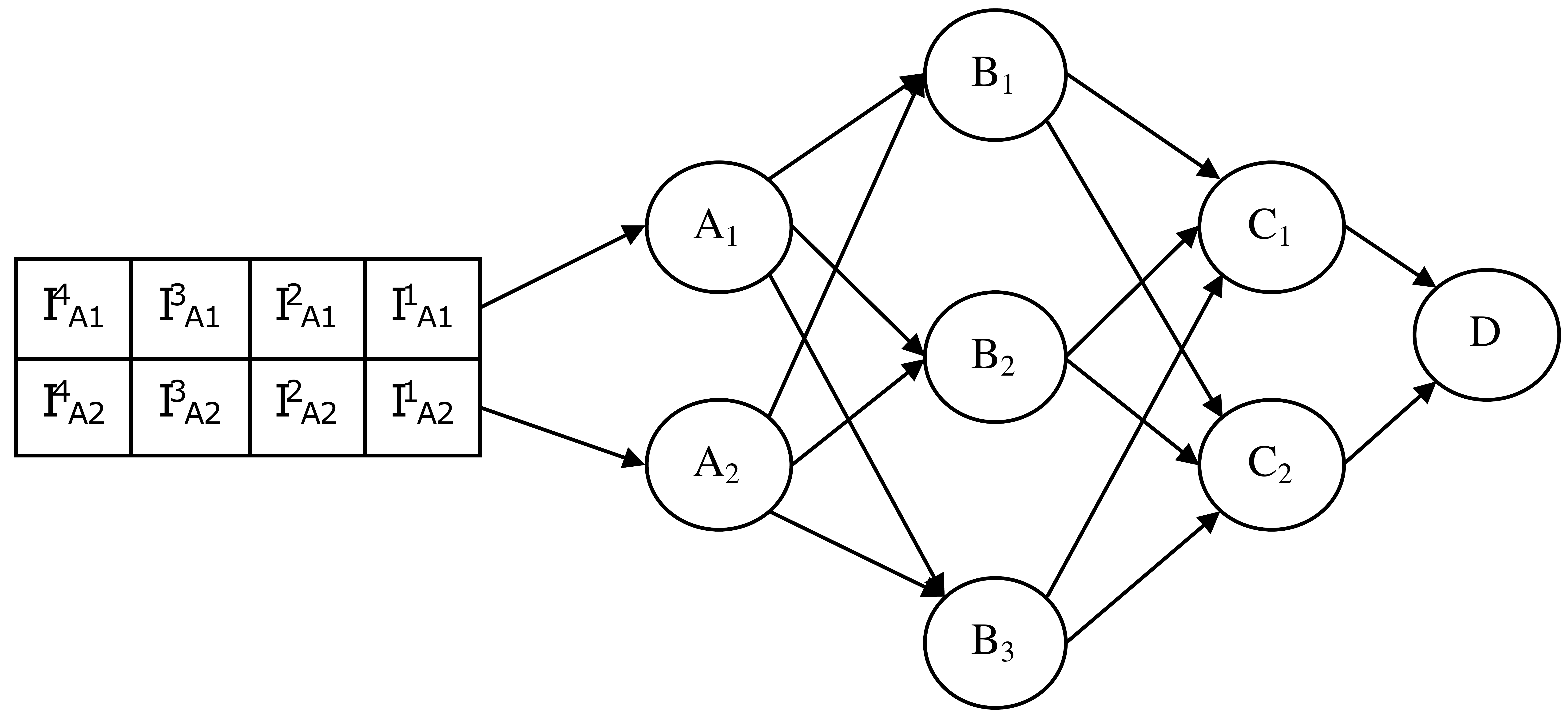}
    \caption{An example of a neural network.}
    \label{fig:dnn}
\end{figure}

In this section, we take an example of a neural network as shown in Figure \ref{fig:dnn} to explain the parallelism. In the example, we assume that the model contains three layers and seven data processing nodes (neurons), i.e., $A_1$, $A_2$, $B_1$, $B_2$, $B_3$, $C_1$, $C_2$, $D$. The arrows represent the data flow among different data processing nodes. The execution of the data processing nodes at each layer can be carried out in parallel, while the execution of different layers should be performed sequentially. The input data contains 4 data points. We assume two/three computing resources owned by two/three users. Each has a part of the input data.

\begin{figure}[htbp]
    \centering
    \includegraphics[width=0.6\textwidth]{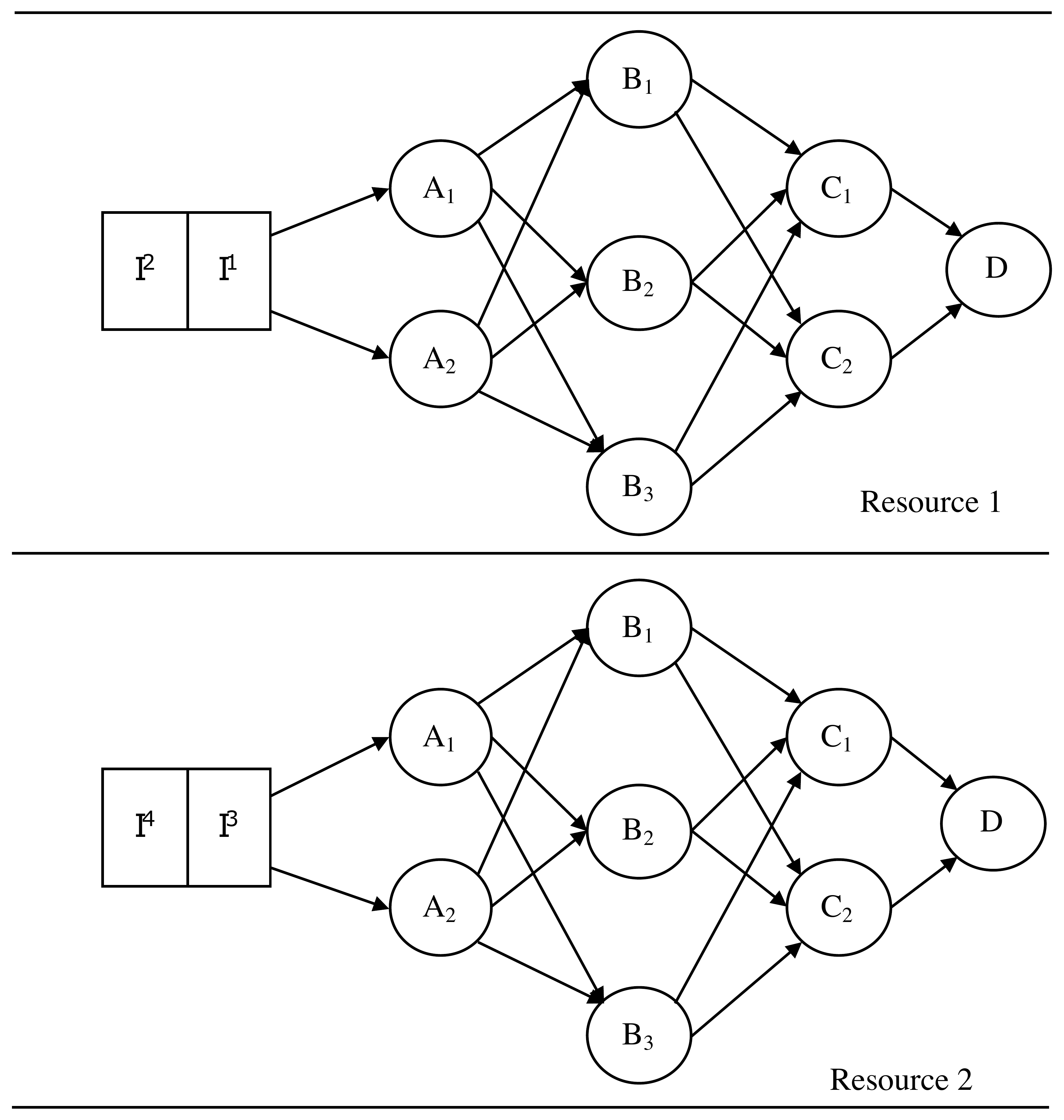}
    \caption{Data parallelism. The forward and backward process of $I^1$ and $I^2$ is performed in computing resource 1, while that of $I^3$ and $I^4$ is performed in computing resource 2 at the same time. Then the model or gradient is transferred to calculate an average model or gradient to be sent to each computing resource for the following training.}
    \label{fig:datap}
\end{figure}

\subsubsection{Data Parallelism}
Data parallelism is realized by having the data processing performed in parallel at different computing resources, with the same model, on different data points. As shown in Figure \ref{fig:datap}, data parallelism is exploited when the ensemble of data points is distributed among different computing resources. During the training process of FL, the training data is not transferred among different computing resources, while the intermediate data, e.g., the models or the gradients $\nabla F_k(x)$ in Formula \ref{eq:gd}, are transferred. 
The data in each computing resource can be Independent and Identically Distributed Data (IID) or non-IID. FL focuses on the non-IID \cite{mcmahan2017communication}, while other distributed machine learning approaches mainly focus on IID data. 
With the data parallelism, the FL is horizontal \cite{yang2019federated}, i.e., the data and the calculation are horizontally distributed among multiple computing resources.
In addition, this parallelism generally corresponds to the cross-device FL \cite{kairouz2019advances}, where a large number of devices (mobiles or edge devices) collaboratively participate in training a single global model to have good accuracy.
When the number of devices is small, e.g., 2-100, and the computing resources are from diverse organizations, this parallelism also corresponds to cross-silo FL \cite{kairouz2019advances}. In addition to the general data-parallel schemes for federated learning, some specific privacy-preserved distributed statistical tricks have been invented for federated sparse models~\cite{bian2017multi,bian2020mp2sda}.

\begin{figure}[htbp]
    \centering
    \includegraphics[width=0.75\textwidth]{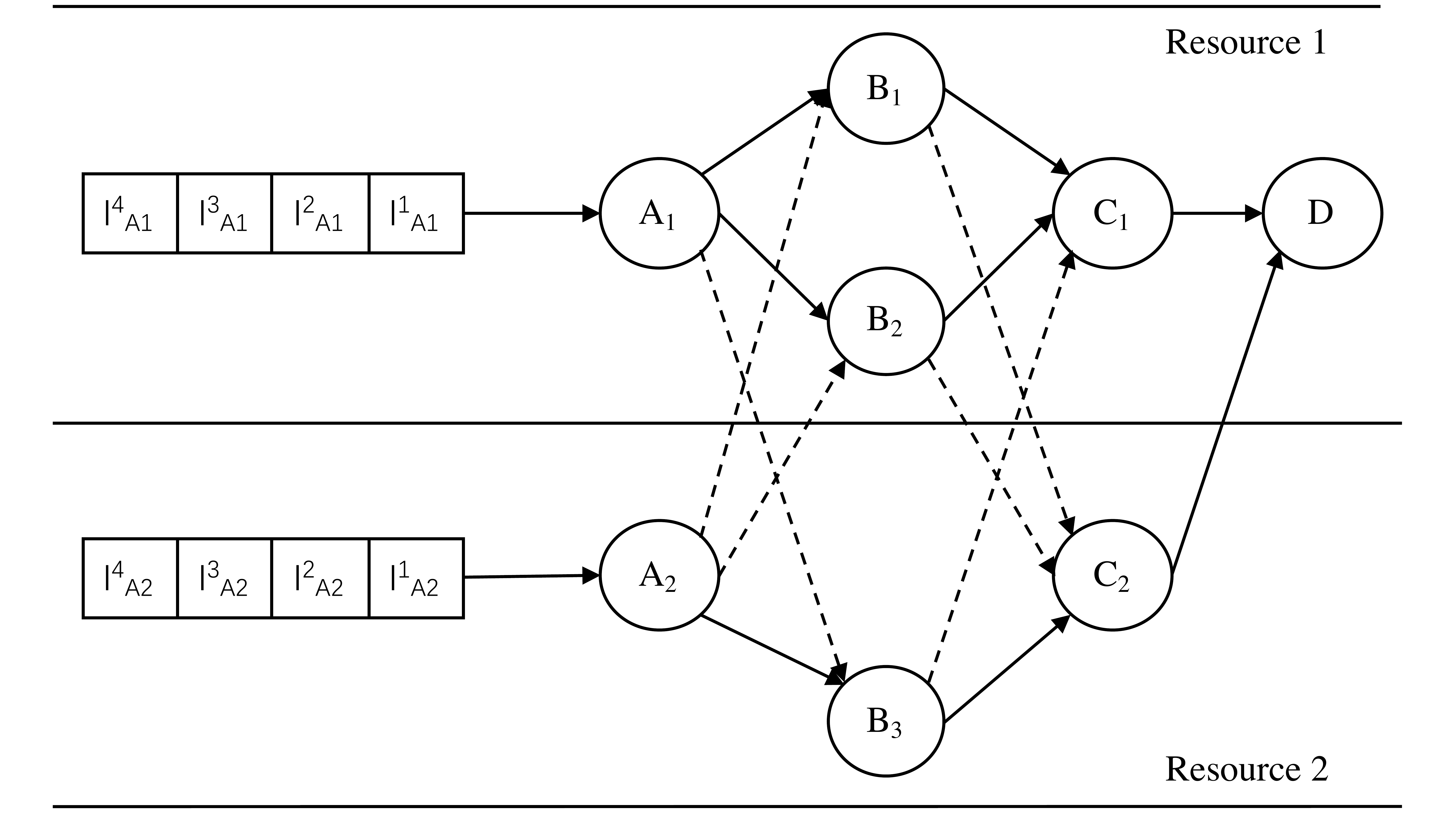}
    \caption{Model parallelism. The dashed arrows represent inter-computing resource communication. For each input data point $I$, different parts, i.e., $I_{A_1}$ and $I_{A_2}$, are distributed at different computing resources.}
    \label{fig:modelp}
\end{figure}

\subsubsection{Model Parallelism}
\label{subsubsec:modelP}
Model parallelism is realized by having independent data processing nodes distributed at different computing resources, so as to process the data points of specific features. 
Two data processing nodes can be either independent, i.e., the execution of any node does not depend on the output of the other one; or dependent, i.e., there is a data dependency between them \cite{liu2015survey}.
As shown in Figure \ref{fig:modelp}, model parallelism is achieved when different parts of each data point are distributed at different computing resources. 
For instance, the data process on Node $A_1$ and that of $A_2$ can be carried out in parallel.
With the model parallelism, vertical FL, where the data points and calculation are vertically distributed among multiple computing resources \cite{hardy2017private, yang2019federated}, is realized.
In this case, the original model needs to be partitioned to be distributed at different computing resources.
Two organizations generally apply this type of FL when each organization owns parts of the features of users and they would like to collaboratively train a model using the data of all the features, which corresponds to cross-silo FL \cite{kairouz2019advances}.
Most studies of vertical federated learning only support two parties (with or without a central coordinator) \cite{zhu2021}. 
For instance, SecureGBM \cite{feng2019securegbm} is proposed to train a tree-based Gradient Boosting Machine (GBM).
In order to support multiple parties, the idea of multi-view learning \cite{xu2013survey} is exploited in a multi-participant, multi-class vertical federated learning framework \cite{feng2020multi}.

\begin{figure}[htbp]
    \centering
    \includegraphics[width=0.7\textwidth]{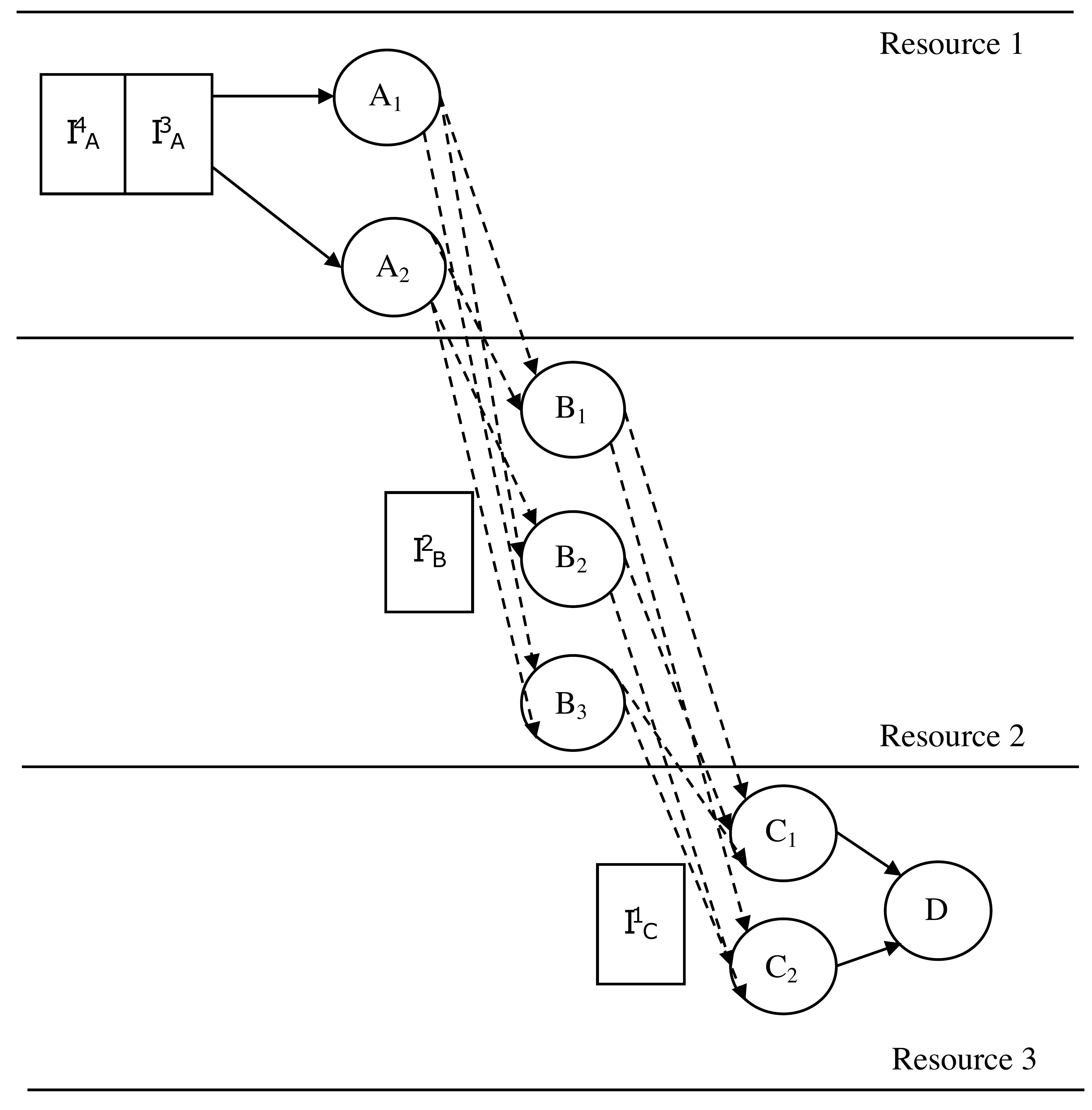}
    \caption{Pipeline parallelism. The dashed arrows represent inter computing resource communication.}
    \label{fig:pipelinep}
\end{figure}

\subsubsection{Pipeline Parallelism}
Pipeline parallelism is realized by having dependent data processing nodes distributed at different computing resources \cite{narayanan2019pipedream, huang2018gpipe}. As shown in Figure \ref{fig:pipelinep}, the data processing nodes are distributed at multiple computing resources. While data point $I^3_A$ is processed in computing resource 1, the outputs of $A_1$ and $A_2$ are processed in computing resource 2, and the outputs of $B_1$, $B_2$, and $B_3$ are processed in computing resource 3. With this type of parallelism, the dependent data processing nodes can process the data in parallel. As this parallelism may incur many inter-computing resource data transfers, it is not widely used for FL. 

\subsection{Aggregation Algorithms}
\label{sebsec:aggre}

With the horizontal FL and data parallelism, aggregation algorithms are used to aggregate the models or gradients generated from the forward and backward propagation in each computing resource. The aggregation algorithms can be either centralized, or hierarchical, and decentralized. The centralized aggregation algorithms generally rely on a centralized server, i.e., a parameter server, to synchronize or schedule the execution of distributed computing resources, while hierarchical aggregation algorithms rely on multiple parameter servers for the model aggregation.
The decentralized aggregation algorithms make each computing resource equally perform the calculation based on a predefined protocol, without relying on a centralized server. \liu{Please refer to \cite{Wang2021Guide} for the details of federated optimization. The characteristics are summarized in Table \ref{tal:aggregation}, which can be used to choose appropriate algorithms in a specific situation.  }

\begin{table}[ht]
\centering
\caption{\liu{Comparison among diverse types of aggregation algorithms. ``Complexity'' represents the complexity to implement the algorithms (``H'' represents high complexity, ``M'' represents medium complexity, and ``L'' represents low complexity). ``Trust'' represents whether the aggregation algorithms require that the data owners trust the centralized server. ``Imbalance'' represents whether the algorithms can address the unbalanced data. ``High-latency'' represents whether the algorithms can support the high-latency model or gradient data transfer. ``Y'' represents that the algorithms support the functionality while ``N'' represents that the algorithms do not support the functionality.}}
\label{tal:aggregation}
\begin{tabular}{|c|c|c|c|c|c|}
\hline
Type & Complexity & Trust & Imbalance & High-latency \\
\hline
Centralized & L & Y & N & N \\
\hline
Hierarchical & M & Y & Y & Y \\
\hline
Decentralized & H & N & N & Y \\
\hline
\end{tabular}
\end{table}

\subsubsection{Centralized Aggregation}
\label{subsubsec:centralized}

\begin{figure}[htbp]
    \centering
    \includegraphics[width=0.7\textwidth]{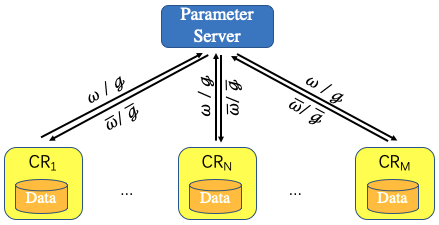}
    \caption{The architecture of centralized aggregation. ``CR'' represents computer resource. $\omega$ represents the local parameters or the weights of the model calculated in each computing resource. $\mathrm{g}$ represents the local gradients in backward propagation in each computing resource. $\overline{\omega}$ represents the global model calculated in the parameter server. $\overline{\mathrm{g}}$ represents the global gradients calculated in the parameter server.}
    \label{fig:centra}
\end{figure}

As shown in Figure \ref{fig:centra}, a single parameter server is used to calculate the average models or gradients sent from multiple computing resources (mobiles).
The weights of the model (model) or the gradients are calculated in each computing resource, which are transferred to a parameter server. 
The parameter server calculates global gradients or global models according to a centralized aggregation algorithm.
The global gradients or global models are transferred to each computing resource for the following computation. 
The update of the model is based on the SGD defined in Formula \ref{eq:gd} in both computing resources, or on the parameter server.

A bunch of centralized aggregation algorithms have been proposed.
Federated Averaging (FedAvg) \cite{mcmahan2017communication} algorithm is introduced as the aggregation method in Google's implementation of an FL system. A centralized server aggregates the machine learning models from selected users. Then, a global model is generated using a weighted sum of each aggregated machine learning model. Afterward, the global model is shared with selected users, and the training process is continued in the computing resource of selected users. 
However, the trained model of FedAvg may be biased towards computing resources with favorable network conditions \cite{li2020federated}. 
While FedAvg is a straightforward approach, some other methods are proposed to address additional problems. 
A Federated Stochastic Block Coordinate Descent (FedBCD) \cite{liu2019communication} algorithm is proposed to reduce the number of communication rounds by enabling multiple local updates before the model communication between a user and the server. In addition, FedBCD also considers the regularization during the training process. The training problem with regularization can be formulated as:
\begin{equation}
\label{eq:regularizer}
    \widehat F = \underset{F_\theta\in\mathcal{H}}{\mathrm{argmin}} \underset{(x,y)\sim\mathcal{D}}{\mathbb{E}} L(y, F_\theta(x)) + \lambda \cdot \gamma(\theta),
\end{equation}
where $F$, $D$, $\mathcal{H}$ are the same as those in Formula \ref{eq:problem}, while $\gamma(\cdot)$ denotes the regularizer and $\lambda$ is the hyper-parameter. The regularization is exploited to improve the generalization capacity of the trained machine learning model. As the fairness among multiple users is important for an FL system, the Stochastic Agnostic Federated Learning (SAFL) \cite{mohri2019agnostic} algorithm and the FedMGDA+ \cite{hu2020fedmgda} algorithm are proposed to achieve fairness during the training process of FL. The fairness represents that the data distribution among multiple users can be equally considered without the influence of unrelated factors. 
Fairness may also refer to two other concepts: (1) A user gets a final model according to the contribution \cite{lyu2020towards}; and/or (2) Uniform accuracy distribution among all the distributed computing resources \cite{li2019fair}, which are out of the scope of this paper.
In addition, while the computing resources may be heterogeneous, FedProx \cite{Li2020} is proposed to tackle the heterogeneity in an FL system. FedProx enables multiple iterations in each computing resource, while minimizing a cost function based on the local loss function and the global model. Furthermore, in order to address permutation of data processing nodes during the training process, Federated Matched Averaging (FedMA) \cite{Wang2020Federated} is proposed. FedMA exploits an existing approach, i.e., BBP-MAP \cite{yurochkin2019bayesian}, to generate a matrix, in order to align the data processing nodes of the models from computing resources and the server.  SCAFFOLD \cite{karimireddy2020scaffold} is proposed to reduce the communication rounds, using stateful variables in the distributed computing resources. Attention-augmented mechanism is exploited in Attentive Federated Aggregation (FedAttOpt) \cite{jiang2020decentralized} to aggregate the knowledge generated from each computing resource (client), based on the contribution of the model from each client. 
When the data distribution is heterogeneous among users, personalization remains an open problem. In order to address this problem, the model can be split into local layers and global layers, which has been proposed in adaptive personalized federated learning (APFL) \cite{deng2020adaptive}, FedPer \cite{arivazhagan2019federated}, and pFedMe \cite{dinh2020personalized}. The local layers are trained with the decentralized data in each computing resource of users, while the global layers are trained in the computing resources of users and the server. However, it is difficult to choose a dataset and its partition among clients to measure the personalization brought by APFL or FedPer, so as to prove the improvement compared with FedAvg.
The attention-augmented mechanism helps reduce the communication rounds.
In addition, knowledge distillation can also be exploited to aggregate the models, while requiring that there is data in the centralized server \cite{He2020Group}.
All these algorithms can handle non-IID data. A comparison among the aforementioned algorithms is proposed in Table \ref{tal:aggre}. 

\begin{table}[ht]
\centering
\caption{Comparison among aggregation algorithms. ``Reg'' represents regularization. Heterogeneity represents that the computing resources are heterogeneous. ``Firness'' represents that the data distribution among multiple users can be equally considered without the influence of unrelated factors. ``Permutation'' refers to the permutation of data processing nodes during the training process.  ``C-E'' represents communication efficient. ``S'' represents that the algorithm supports the functionality, while ``N'' represents that the algorithm does not have support. }
\label{tal:aggre}
\begin{tabular}{|c|c|c|c|c|c|}
\hline
Algorithm & Reg & Fairness & Heterogeneity & Permutation & C-E \\
\hline
FedAvg & N & N & N & N & N \\
\hline
FedBCD & S & N & N & N & S \\
\hline
SAFL & N & S & N & N & N \\
\hline
FedMGDA+ & N & S & N & N & S \\
\hline
FedProx & N & N & S & N & S \\
\hline
FedMA & N & N & N & S & S \\
\hline
SCAFFOLD& N & N & N & N & S \\
\hline
FedAttOpt& N & N & N & N & S \\
\hline
\end{tabular}
\end{table}

\subsubsection{Hierarchical Aggregation}

\begin{figure}[htbp]
    \centering
    \includegraphics[width=0.95\textwidth]{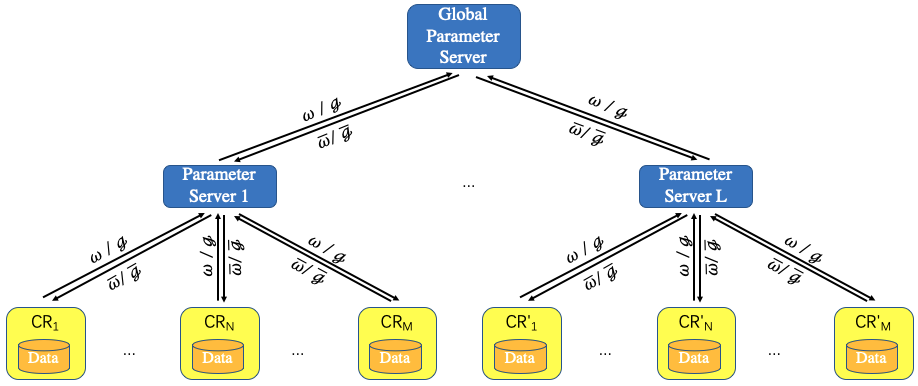}
    \caption{The architecture of hierarchical aggregation. ``CR'' represents computer resource. $\omega$ represents the local parameters or the weights of the model calculated in each computing resource. $\mathrm{g}$ represents the local gradients in backward propagation in each computing resource. $\overline{\omega}$ represents the region or global model calculated in each parameter server. $\overline{\mathrm{g}}$ represents the region or global gradients calculated in each parameter server. The region model or gradients are calculated by a region parameter server, while the global model or gradients are calculated by a global parameter server.}
    \label{fig:hier}
\end{figure}

As shown in Figure \ref{fig:hier}, a hierarchical architecture is also exploited using multiple parameter servers. 
A two-layer hierarchical architecture is proposed to reduce the time to transfer models between a parameter server and computing resources \cite{abad2020hierarchical}. The hierarchical architecture uses a global parameter server (GPS) and multiple region parameter servers. Each region parameter server (RPS) is implemented in a cell base station where the computing resources (mobiles) can be connected, with low latency. A hierarchical algorithm, i.e., Hierarchical Federated Learning (HFL) is deployed to realize the model aggregation. Within each iteration of HFL, each RPS calculates an average model using the models of the computing resources within its cluster. It sends the averaged model to the GPS, and it receives a global averaged model at every certain iteration. Afterward, it broadcasts the averaged model to all its computing resources. Some other algorithms, e.g., HierFAVG \cite{liu2020client}, HFEL \cite{luo2020hfel}, and LanFL \cite{yuan2020hierarchical}, are similar to HFL, while the SPS is an edge or Local-Area Network (LAN) parameter server and the MPS is a parameter server implemented on the cloud or a Wide-Area Network (WAN). These algorithms take advantage of hierarchical architecture to reduce high-latency model or gradient data transfer, so as to accelerate the training process. In addition, by well-clustering the computing resources to groups, the hierarchical architecture is also exploited to address unbalanced data distributed among multiple computing resources \cite{briggs2020federated,mhaisen2021optimal}, or to address data privacy \cite{wainakh2020enhancing}.

\subsubsection{Decentralized Aggregation}
\label{subsubsec:decentralized}

\begin{figure}[htbp]
    \centering
    \includegraphics[width=0.8\textwidth]{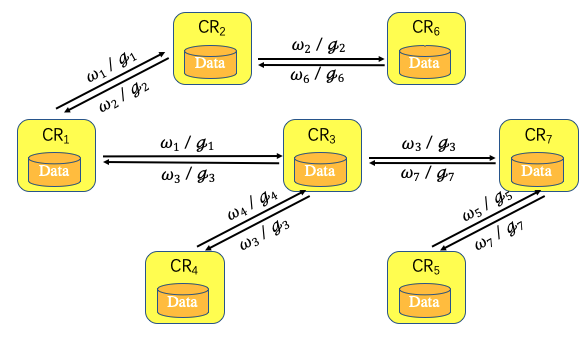}
    \caption{The architecture of decentralized aggregation. ``CR'' represents computer resource. $\omega$ represents the local parameters, or the weights of the model calculated in each computing resource. $\mathrm{g}$ represents the local gradients in backward propagation in each computing resource. When two computing resources are neighbors, they can communicate with each other.}
    \label{fig:decentralized}
\end{figure}

While collaboratively training a machine learning model with a decentralized aggregation algorithm, the computing resources can be organized with a {\em connected} topology and can communicate with a peer-to-peer manner, as shown in Figure \ref{fig:decentralized}. The degree and connectivity of the topology affects the communication efficiency and the convergence rate of the aggregation algorithm. 
For a given topology, we define $w_{i,j}$, the weight to scale information flowing from node $j$ to node $i$, as follows\vspace{-2mm}
\begin{align}\label{wij}
w_{ij}
\begin{cases}
> 0 & \mbox{if node $j$ is connected to node $i$, or $i=j$;} \\
= 0 & \mbox{otherwise.}\vspace{-2mm}
\end{cases}
\end{align}
We further define the {\bf topology matrix} $W = [w_{ij}]_{i,j=0}^{n-1} \in \mathbb{R}^{n\times n}$ as the matrix to represent the topology. 
In the remainder of this paper, 
we assume that $W$ satisfies 
$W\mathds{1} = \mathds{1}$ 
and $\mathds{1}^T W = \mathds{1}^T$, 
i.e., both the row sum and the column sum of $W$ are equal to $1$, 
so as to guarantee that the neighborhood averaging will asymptotically approach the global averaging \cite{chatterjee1977towards, seneta2006non, sayed2014adaptation}. 
When a computing resource $j$ is directly connected to computing resource $i$, i.e., $w_{i,j} \neq 0$, computing resource $j$ is the neighbor of computing resource $i$.
\liu{Please note that the weight $w_{i,j}$ denotes the confidence node $i$ has in the information it receives from node $j$ \cite{lalitha2019peer}, which is different from the bandwidth or data transfer capacity in a network.}
The centralized aggregation algorithm is a special type of decentralized aggregation with a star topology while only the centralized server communicates with its neighbors.
\liu{A well designed topology, e.g., an exponential graph \cite{Assran2019Stochastic}, can improve the convergence rate, which accelerates the training speed.}

With the decentralized SGD (D-SGD), each computing resource maintains a local copy of 
the global model parameters, and it updates the local copy using the models of its neighbors. 
According to the order to conduct neighborhood averaging and gradient descent, D-SGD has two common types of realizations: Average-With-Communication (AWC) \cite{lian2017can, lalitha2019peer} and Average-Before-Communication (ABC) \cite{chen2012diffusion, wang2019matcha}.
AWC can overlap communication and gradient computation, while ABC needs to sequentially calculate and communicate the gradient or model. 
However, ABC is robust \cite{sayed2013diffusion}, and it converges fast in terms of iterations by exploiting its large learning rate.

In addition, the decentralized aggregation algorithms can be classified to Full Communication (FC) \cite{lian2017can} and Partial Communication (PC) \cite{wang2019matcha, vanhaesebrouck2017decentralized} according to the number of neighbors. Within the iterations of FC, each computing resource calculates an averaged model or gradient, based on all the models or gradients of the last version from all its neighbors. However, within the iterations of PC, each computing resource calculates an averaged model or gradient based on one or multiple chosen neighbors. With PC, the selection of the neighbors can be based on a gossip algorithm \cite{Hu2019Decentralized}. For instance, a random neighbor can be selected \cite{vanhaesebrouck2017decentralized}; the neighbors that provide benign models are selected to avoid attack \cite{munoz2019byzantine}.

\section{Data Manipulation}
\label{sec:security}

At the infrastructure layer of an FL system, there are three types of data manipulation: data security mechanisms, data transfer, and distributed data processing within the distributed execution module. 
We first present the techniques for the distributed execution in an FL system. 
Then, we present the techniques for data transfer during the training process of an FL system.
Finally, as data security is of much importance to an FL system \cite{mothukuri2021survey}, we present the techniques to protect the data security. 

\subsection{Distributed Data Processing}

While the bandwidth within a single data center is high, e.g., InfiniBand, the High Performance Computing (HPC) libraries, e.g., Message Passing Interface (MPI) \cite{gropp1999using} or NVIDIA Collective Communications Library (NCCL) \cite{awan2018optimized}, are widely exploited for distributed data processing \cite{vishnu2016distributed}. With MPI or NCCL, the gradients or models in each computing resource can be easily calculated using ring-AllReduce algorithm \cite{ring-allreduce}. However, one of the drawbacks of the HPC libraries is that they lack support for fault-tolerance, as the HPC libraries are designed for high performance servers with high quality networks. When any computing resource within the network becomes unavailable, the distributed training process may be broken.

However, as an FL system is generally implemented for the collaboration of large amounts of mobile device users or different organizations, the network connection among computing resources is of moderate quality, i.e., the bandwidth is not as good as that within a single data center, and the latency is high.
For instance, the Internet upload speed is typically much slower than the download speed \cite{konevcny2016federated}. Also, some users with unstable wireless communication channels may consequently drop out due to disconnection from the Internet \cite{lim2020federated}.
In this environment, the connection between computing resources and parameter servers has a high possibility of becoming disabled. Thus,  Remote Procedure Call (RPC) frameworks are widely exploited, as this kind of framework can ignore the disconnected computing resources and continue the distributed training of an FL system \cite{beutel2020flower}, e.g., PaddleFL \cite{PaddleFL}, PySyft \cite{Pysyft}, or TensorflowFL \cite{TFF}.

\subsection{Data Transfer}

As the network connection is of moderate quality, the data transfer module mainly focuses on data compression to transfer intermediate data, e.g., gradients or models. 
Sketched updates are proposed for gradient compression to accelerate the data transfer during the distributed training within a single data center \cite{jiang2018sketchml, karimireddy2019error, ivkin2019communication, spring2019compressing}.
With the data parallelism and centralized aggregation algorithm, before sending the intermediate data, the intermediate data can be sketched with subsampling \cite{konevcny2016federated}, quantization \cite{konevcny2016federated, haddadpour2020federated, shlezinger2020uveqfed ,shlezinger2020federated, sun2020adaptive, xu2020ternary}, sparsification \cite{sun2020adaptive, Malekijoo2021FEDZIP}, or projection to lower dimensional spaces \cite{rothchild2020fetchsgd}, in each computing resource, in order to reduce the cost to transfer data.
Subsampling refers to transferring only a random subset of the intermediate data \cite{konevcny2016federated}.
Quantization methods encode each value using a fixed number of bits, so as to reduce the length of gradients or models \cite{konevcny2016federated}.
With the sparsification approach, only selected parts of the intermediate data are transferred, while the selection is based on a threshold, e.g., the gradients larger than a threshold are selected \cite{sun2020adaptive}. Then, when the intermediate data is received in the server, they are decompressed to be aggregated according to the aggregation algorithms presented in Section \ref{subsubsec:centralized}. The convergence of the quantization approach is analyzed in \cite{haddadpour2020federated}, which shows that this approach can also provide good convergence rates \cite{haddadpour2020federated}. In addition, irrelevant intermediate data can be precluded to be transferred to the server, in order to substantially reduce the communication overhead \cite{luping2019cmfl}.

\subsection{Data Security}

Data security is of much importance for data processing. The problem of data security is related to significant financial \cite{UberLoss,GoogleLoss} and reputational \cite{FacebookLoss} losses. For instance, Uber had to pay \$148,000,000 to settle the investigation incurred by a breach of 600,000 drivers’ personal information in 2016 \cite{Uber}. 
Data security mainly includes two aspects, i.e., data privacy and model security.
Data privacy refers to the protection of raw data to avoid raw data information leakage during or after the distributed training of FL systems.
Model security refers to the protection of the security of trained models, in order to avoid wrong output based on the trained models incurred by malicious attacks.
In this section, we first present the techniques to protect data privacy. Then we present the defense methods for model security.

\subsubsection{Data Privacy}

The techniques to protect data privacy consist of three types: Trusted Execution Environment (TEE), encryption, Differential Privacy (DP), and anti-Generative Adversarial Network (GAN) methods. These techniques can be combined in FL systems, e.g., the combination of DP and TEE in \cite{hao2019towards}, the combination of encryption and DP \cite{zhang2020batchcrypt}, and the combination of DP and anti-GAN \cite{triastcyn2020federated}.

A TEE is an environment where the execution is secured and no information can be leaked to unauthorized users. 
Intel SGX technique \cite{mckeen2013innovative} has been first proposed as a secure environment while providing a set of security-related instruction codes built within Intel Central Processing Units (CPUs). 
Then, the implementation of machine learning models has been carried out in the TEE, i.e., Intel SGX, in order to enable collaborative data analysis based on machine learning algorithms while providing a security guarantee \cite{ohrimenko2016oblivious}. Afterwards, the TEE has been exploited in FL systems, in order to protect the privacy of data in two ways. The first way is to put the entire training process in the TEE of each distributed computing resource to protect the data privacy during the distributed training \cite{mo2019efficient, chen2020training}. The second way is to use TEE to check a small part of the distributed training, while exploiting insecure computing resources, e.g., GPUs, to reduce the training time \cite{zhang2020enabling}. 

As an encryption technique, homomorphic encryption has been used to ensure the data privacy for FL systems \cite{hao2019towards, feng2019securegbm}.
Homomorphic encryption \cite{yi2014homomorphic} allows specific types of computations to be carried out on encrypted input data, and to generate an encrypted result, which matches the result of the same computations on the decrypted input data. 
\liu{Two main branches of homomorphic encryption exist, i.e., fully homomorphic encryption and partially homomorphic encryption. 
The fully homomorphic encryption supports both addition and multiplication on ciphertext, while partially homomorphic encryption only supports either an addition or a multiplication operation on ciphertext, which corresponds to less computational flexibility and better runtime efficiency. 
Both the fully and partially homomorphic encryptions can be exploited with the horizontal and vertical federated learning.}
As sharing gradients also leaks the information of training data \liu{in horizontal federated learning} \cite{zhao2020idlg, li2019quantification, geiping2020inverting}, it is of much importance to protect the privacy of the intermediate data.
Thus, the intermediate data can be encrypted using a homomorphic encryption algorithm before being sent to a parameter server \cite{lyu2020towards, mandal2019privfl}. In this way, the intermediate data remain encrypted during the aggregation process while only the computing resource can decrypt the encrypted intermediate data. 
Even if the transferred encrypted intermediate data is leaked, the information of gradients or models remains safe, and the privacy of the training data is ensured.
\liu{In addition, partial homomorphic encryption, e.g., Paillier \cite{paillier1999public}, is exploited in vertical federated learning \cite{ccatak2018cpp, ccatak2015secure}.}
However, the homomorphic encryption incurs significant costs in computation and communication during distributed training \cite{zhang2020batchcrypt}. In order to reduce the overhead of homomorphic encryption, a set of quantized gradients are encrypted \cite{zhang2020batchcrypt}. 

Differential Privacy (DP) protects the data privacy by adding artificial noise to a small part of raw data, while ensuring that the modification does not substantially affect the performance of the machine learning models \cite{wei2020federated, dwork2008differential, abadi2016deep, geyer2017differentially}. 
DP is widely used in FL systems as the first step to process the raw data, and the output is the training data to be used for the distributed training \cite{wei2020federated, phan2020scalable, sabater2020distributed, liang2020exploring, liu2020fedsel, katevas2020policy}. With more added noise, the privacy is better protected, i.e., there is less possibility to leak raw data information, while it takes more time to converge for the machine learning models \cite{wei2020federated}. A trade-off between the privacy protection and the convergence performance can be made by selecting a certain number of distributed resources \cite{wei2020federated, seif2020wireless, truex2019hybrid}. However, DP may not be able to ensure the data privacy under certain attacks, e.g., Generative Adversarial Network (GAN) attacks \cite{hitaj2017deep}.

A well-trained machine learning model can leak information about the training data based on the intermediate data, e.g., gradients \cite{hitaj2017deep, ateniese2015hacking, melis2019exploiting}. GANs can be used to generate data similar to the training data based on a well-trained machine learning model \cite{Goodfellow2015Explaining} in either a parameter server \cite{wang2019beyond} or a distributed computing resource \cite{hitaj2017deep}. 
The adversary can reconstruct other participating clients’ private data, even if it has no knowledge of the label information using the GANs.
Thus, during the distributed training process of FL systems, a malicious user can exploit GANs to infer the training data of other users. DP can be used to prevent the GAN-based attack \cite{hitaj2017deep, triastcyn2020federated}. In addition, fake training data can be generated based on a GAN and original raw data, which is then used during the distributed training process to prevent the GAN-based attack \cite{luo2020exploiting}.

\subsubsection{Model Security}

We mainly focus on poisoning attacks in this section. 
The objective of poisoning attacks is to reduce the accuracy of machine learning models using artificially designed data, i.e., data poisoning, or models, i.e., model poisoning, in one or several distributed computing resources during the model aggregation process (see details in in Section \ref{subsubsec:centralized}). There are two ways to carry out poisoning attacks, i.e., data poisoning and model poisoning. 

Data poisoning can be realized by modifying the features \cite{fung2018mitigating} or the labels \cite{tolpegin2020data} of the input data. For instance, malicious users can modify the data points of a certain class $C$ to other classes, and they can then use the modified data points to participate in the distributed training. 
The modification of the labels is denoted by the label flipping attack.
As a result, the accuracy of the trained model has low accuracy in terms of Class $C$ \cite{tolpegin2020data}. 
Model poisoning refers to the attacks in which the updated intermediate data, e.g., gradients or models, are poisoned before being sent to a parameter server in order to reduce the accuracy of the trained model \cite{chen2020backdoor, sun2019can}. The goal of the model poisoning is to reduce the performance of the trained model on targeted tasks or classes, while the performance of the model remains unchanged in terms of other tasks or classes \cite{sun2019can}. 
Data poisoning eventually realizes the model poisoning, as it enables some computing resources to update poisoned intermediate data based on the calculation of poisoned training data \cite{fung2018mitigating}. 
However, model poisoning can be more powerful than data poisoning, as model poisoning directly influences the weights of the models and trains in a way that benefits the attack \cite{bagdasaryan2020backdoor}.
Both the data poisoning and the model poisoning rely on the backdoor attacks to modify the training data or the intermediate data \cite{sun2019can, chen2020backdoor, fung2018mitigating}. Backdoor attacks are performed by embedding the hidden instructions into machine learning models, so that the infected model performs well on benign testing samples when the backdoor is not activated, while its prediction will be changed to the attacker-specified target label when the backdoor is activated by the attacker \cite{li2020backdoor}. 

In order to defend against these data attacks or model attacks, the malicious users should be identified by analyzing the updated intermediate data using dimensionality reduction methods, e.g., Principal Component Analysis (PCA) \cite{tolpegin2020data}, anomaly detection \cite{li2019abnormal, lin2019free}, or interpretability techniques \cite{bhagoji2019analyzing}.
In addition, the model poisoning can be incurred by Byzantine failures of certain distributed computing resources \cite{fang2020local}. With Byzantine failures, some computing resources (bad users) are manipulated by attackers during the distributed training process, which significantly degrades the performance of the global model in terms of test error \cite{fang2020local}. 
In order to make the training process robust against the Byzantine failures, the bad users can be identified by analyzing the updated intermediate data using a hidden Markov model \cite{eddy2004hidden, munoz2019byzantine} or via secure aggregation protocols \cite{he2020secure}. 

\section{Federated Learning Frameworks}
\label{sec:frameworks}

FL systems are widely applied in diverse domains, e.g., 
mobile service, healthcare \cite{xu2020federated}, and finance \cite{li2019survey}.
An FL system generally exploits an FL framework, which is deployed on distributed resources.
In this section, we present four widely used FL frameworks: PaddleFL \cite{PaddleFL}, TensorFlowFederated \cite{TFF}, FATE \cite{FATE}, and PySyft \cite{Pysyft}.

\subsection{PaddleFL}

PaddleFL is an open source federated learning framework based on PaddlePaddle \cite{Ma2019}, which is supported by Baidu. At the presentation layer, PaddleFL provides a textual UI for the interaction between users and the FL system. At the User Services layer, PaddleFL provides the log and monitoring supports, and it can leverage the interpretability module \cite{PaddleInterpretability} of PaddlePaddle in the future. At the FL training layer, PaddleFL can realize data parallelism (horizontal FL) and model parallelism (vertical FL). It supports multiple aggregation algorithms, e.g., FedAvg, and fault-tolerance. At the infrastructure layer, PaddleFL exploits RPC for the distributed execution. PaddleFL exploits DP to protect the data security. PaddleFL is widely used in multiple domains, e.g., Natural Language Processing (NLP), Computing Vision (CV) \cite{liu2020fedvision}, and recommendation.

\subsection{TensorFlowFederated}

TensorFlow Federated (TFF) \cite{TFF} is an open-source framework for federated learning on decentralized data, which is supported by Google. TFF also provides a textual UI through Python. TFF supports the monitoring and log functionality at the user service layer. TFF supports data parallelism (horizontal FL), multiple aggregation algorithms, and fault-tolerance of mobile devices. TFF exploits RPC for the distributed execution and DP for the protection of data privacy.
TFF enables Android mobile users to predict the next word while using the keyboard on their mobile phones \cite{mcmahan2017communication, McMahan2018}. 

\subsection{FATE}

FATE \cite{FATE} is an open-source FL framework supported by WeBank. 
FATE provides both a graphical and textual UI. 
FATE can support the monitoring of distributed training through a web portal. 
FATE takes advantage of database management systems (DBMS) to track the execution status.
FATE can enable horizontal (data parallelism), vertical (model parallelism), and hybrid federated learning. 
FATE exploits both the DP and HE to protect the data privacy. 
In addition, FATE exploits RPC to perform the distributed execution.

\subsection{PySyft}

PySyft \cite{ryffel2018generic} is an open-source FL framework based on the PyTorch framework \cite{Pytorch}.
PySyft is written in Python and provides a textual UI based on Python.
PySyft mainly supports the data parallelism and model parallelism based on an aggregator or orchestrating server. The aggregator or orchestrating server sends a part of the model to participating clients to process local data and gets results for federated averaging. PySyft exploits DP and encryption techniques to protect the data security. PySyft exploits multiple communication protocols for distributed execution, e.g., RPC, websocket \cite{fette2011websocket} etc.

\subsection{\liu{Concluding Remarks}}
\liu{
Diverse FL frameworks exist while each has its advantage. We summarize the characteristics of each framework in Table \ref{tal:FLFramework}, so as to help select a proper framework for use. From the table, we can see that all the frameworks implement the centralized aggregation algorithms, while employing DP and HE for the data security. PaddleFL can exploit Paddle to realize data, model, and pipeline parallelism. FATE and TFF are based on Tensorflow as the engine, while FATE can provide Web portal UI, which is convenient for novices. PySyft is compatible with PyTorch, which can easily handle the PyTorch-based tasks, while PaddleFL is compatible with Paddle, which can easily deal with rich pre-trained models published in PaddleHub \cite{PaddleHub}.}

\begin{table}[ht]
\centering
\caption{\liu{Comparison among diverse frameworks. ``Aggregation'' represents the type of aggregation algorithms. ``Textual'' represents the textual UI, while ``Web'' represents Web portal.}}
\label{tal:FLFramework}
\begin{tabular}{|c|c|c|c|c|c|}
\hline
Framework & Engine & Aggregation & UI & Parallelism & Security \\
\hline
PaddleFL & Paddle & Centralized & textual & Data/Model/Pipeline & DP/HE \\
\hline
TFF & TensorFlow & Centralized & textual & Data/Model & DP/HE \\
\hline
FATE & TensorFlow & Centralized & Web & Data/Model & DP/HE \\
\hline
PySyft & PyTorch & Centralized & textual & Data & DP/HE \\
\hline
\end{tabular}
\end{table}

\liu{Table \ref{tal:FrameworkSupport} represents the support of diverse types of FL in terms of data distribution. All the frameworks support horizontal FL, while vertical FL is supported by three frameworks except TFF. PySyft cannot directly support the vertical FL, while PyVertical \cite{Romanini2021}, which is built upon PySyft, can be used to support vertical FL with the compatibility of PyTorch models. The hybrid FL is only supported by Paddle and FATE. In addition, all the frameworks support the execution with GPU. In practice, although PaddleFL may correspond to slightly longer time, the accuracy of the trained model can be higher that of TFF and FATE, while PySyft may generate ``out of memory'' errors \cite{Kholod2021}.}

\begin{table}[ht]
\centering
\caption{\liu{Comparison among diverse frameworks for the support of diverse FL types, e.g., horizontal FL, vertical FL, and hybrid FL, and GPU. $\|$\checkmark$\|$ represents that the support is not realized by itself but a close one.}}
\label{tal:FrameworkSupport}
\begin{tabular}{|c|c|c|c|c|c|}
\hline
\multicolumn{2}{|c|}{} & PaddleFL & TFF & FATE & PySyft \\
\hline
\multirow{3}{*}{Types} & Horizontal & \checkmark & \checkmark & \checkmark & \checkmark \\
\cline{2-6}
& Vertical & \checkmark & \xmark & \checkmark & $\|$\checkmark$\|$ \\
\cline{2-6}
& Hybrid & \checkmark & \xmark & \checkmark & \xmark \\
\hline
\multicolumn{2}{|c|}{GPU} & \checkmark & \checkmark & \checkmark & \checkmark \\
\hline
\end{tabular}
\end{table}

\section{Research Directions}
\label{sec:future}

Although much work has been done on the FL systems, there remain
some limitations, e.g., interpretability of FL, decentralized aggregation, FL on graphs,  benchmarks of FL systems, and applications to distributed intelligent systems. This section
discusses the limitations of the existing frameworks and proposes new research directions.

\subsection{Benchmarks}

Several datasets exist for experiments on FL systems. For instance, Federated Extended MNIST (FEMNIST) \cite{caldas2018leaf} is built by partitioning the data in Extended MNIST \cite{cohen2017emnist} based on each writer.
Shakespeare \cite{mcmahan2017communication} is built from The Complete Works of William Shakespeare \cite{shakespeare2007complete} based on each speaking role. 
Both of these datasets can be used for horizontal FL. However, no public datasets exist for vertical FL or transfer FL. In addition, no open decentralized IID or non-IID distribution of popular datasets, e.g., ImageNet \cite{deng2009imagenet}, exist for FL systems. 

\subsection{Interpretability}

Deep neural networks have excellent performance in various areas, while it is often difficult to understand the results of deep neural network models, especially within FL systems.
Shapley values have been used to provide the interpretability \cite{wang2019interpret}, while it focuses on vertical FL. When multiple users collaboratively train an FL model, it remains an open problem to evaluate the contributions of each user, which helps provide evidence for the incentive of each user. 
The primary incentive for clients to participate in federated learning is obtaining better models \cite{kulkarni2020survey}, while the benefit of participating in federated learning for clients who have sufficient private data to train accurate local models is disputable. 
Interpretability can help understand the contributions of each user and provide an objective opinion on the incentive strategy within an FL system.
In addition, the interpretability helps domain experts to understand the relationship between data and the final trained model in critical domains, e.g., healthcare and finance.
However, the interpretability within FL systems remains an open problem.

\subsection{Decentralized Aggregation}

Current aggregation algorithms of FL systems focus on the full connection or star connection topology, while other topologies, e.g., dynamic exponential-2 graph, may help accelerate the distributed training with FL systems \cite{lian2017can}. \liu{In addition, well-known graph algorithms, e.g., graph partitioning algorithms, and ad-hoc policies can be exploited to help better distribute computing resources with the topology defined in Section \ref{subsubsec:decentralized} in order to improve the efficiency of FL systems. } While the peer-to-peer communication enables the FL with an arbitrary topology matrix, the data security under diverse attacks, e.g., data or model poisoning, GAN-based attacks, remain open problems and deserve further investigation. 

\subsection{Federated Learning on Graphs}

Graphs or graph neural networks (GNN) \cite{velivckovic2017graph} have gained increasing popularity in multiple domains, e.g., social network, knowledge graph, and recommender system. FL frameworks for graphs, i.e., GraphFL \cite{wang2020graphfl}, and GNN, i.e., SGNN \cite{mei2019sgnn}, have been proposed to train a model with decentralized graphs. However, the data security of FL on graphs remains an open problem. \liu{In addition, while a multimodal knowledge graph could not only contain text but also images or other type of data \cite{zhao2021multimodal}, it is worth further exploration to efficiently support the multimodel knowledge graph construction within an FL system \cite{lin2020improving}.}

\subsection{\liu{Imbalanced Data}}
\liu{
Although FL focuses on the non-IID data, the real-world decentralized data usually exhibit an imbalanced distribution \cite{He2009Imbalanced, wu2021adversarial}. While the imbalanced data exist in multiple areas, such as computer vision \cite{oksuz2020imbalance}, bioinformatics, and biomedicine \cite{zhang2017feature}, learning from such data requires special attention upon data sampling \cite{zhang2017feature, zhang2021empirical}, data augmentation \cite{oksuz2020imbalance}, and loss function designs \cite{wang2021addressing}. The imbalanced data is related to diverse tasks, e.g., two-class or multi-class classification \cite{bi2018empirical, zhang2019multi}. However, an optimized approach can be proposed to address the imbalanced data within FL systems.}

\subsection{Applications to Distributed Intelligent Systems}
Machine learning algorithms have been widely used to boost the performance of intelligent systems, while FL systems could further enhance intelligent systems~\cite{liu2020two} in distributed computing environments \cite{liu2016multi, pineda2016managing} with privacy and security ensured. An intelligent system is a group of machines that has the capacity to gather data, analyze the data, and respond to other systems or the world around. 
With FL systems, the distributed data can be exploited to generate models of high performance so as to produce smart responses. 

\section{Conclusion}
\label{sec:con}

In this paper, we discussed the current state of the art of FL systems, including the functional architecture of FL systems, distributed training, and data manipulation. 

First, we presented an overview of FL systems. In particular, we introduced the life cycle of FL models, including four phases. Then, we presented the four-layer functional architecture of FL systems, including presentation, user services, FL training, and infrastructure, and we presented each layer in detail. 

Second, we detailed the distributed training with two parts, i.e., parallelism and aggregation algorithms. We presented three types of parallelism, including data parallelism, model parallelism, and pipeline parallelism. We associate each parallelism to a corresponding type of FL. For instance, data parallelism is associated with the horizontal FL, which corresponds to cross-device or cross-silo FL. Model parallelism is related to vertical FL and cross-silo FL. We presented the features of different aggregation algorithms in three types: centralized aggregation, hierarchical aggregation, and decentralized aggregation. 

Third, we presented the techniques for data manipulation within FL systems. We showed that FL systems prefer RPC for the distributed execution, to handle the fault-tolerance because of moderate network connection. Intermediate data are sketched in order to compress the data, so as to reduce the data communication time. In addition, we presented the data privacy and model security attacks and corresponding defense techniques, e.g., DP, HE, TEE, and the analysis of updated intermediate data for malicious user identification. 

We mainly introduced four FL systems: PaddleFL, TensorFlowFederated, FATE, and PySyft. The current solutions primarily focus on the horizontal FL. And we identified five research directions that deserve further investigation: benchmarks, interpretability, decentralized aggregation, FL on graphs, imbalanced data, and the applications of FL systems to distributed intelligent systems.

\bibliographystyle{plain}
\bibliography{references}

\begin{thebibliography}{100}

\bibitem{CCPA}
California consumer privacy act home page.
\newblock \url{https://www.caprivacy.org/}.
\newblock Online; accessed 14/02/2021.

\bibitem{abad2020hierarchical}
M~Salehi~Heydar Abad, Emre Ozfatura, Deniz Gunduz, and Ozgur Ercetin.
\newblock Hierarchical federated learning across heterogeneous cellular
  networks.
\newblock In {\em IEEE Int. Conf. on Acoustics, Speech and Signal Processing
  (ICASSP)}, pages 8866--8870, 2020.

\bibitem{abadi2016deep}
Martin Abadi, Andy Chu, Ian Goodfellow, H~Brendan McMahan, Ilya Mironov, Kunal
  Talwar, and Li~Zhang.
\newblock Deep learning with differential privacy.
\newblock In {\em ACM SIGSAC conf. on computer and communications security},
  pages 308--318, 2016.

\bibitem{abou2019co}
Zakaria Abou El~Houda, Abdelhakim Hafid, and Lyes Khoukhi.
\newblock Co-iot: a collaborative ddos mitigation scheme in iot environment
  based on blockchain using sdn.
\newblock In {\em IEEE Global Communications Conference (GLOBECOM)}, pages
  1--6, 2019.

\bibitem{aono2017privacy}
Yoshinori Aono, Takuya Hayashi, Lihua Wang, Shiho Moriai, et~al.
\newblock Privacy-preserving deep learning via additively homomorphic
  encryption.
\newblock {\em IEEE Transactions on Information Forensics and Security},
  13(5):1333--1345, 2017.

\bibitem{arivazhagan2019federated}
Manoj~Ghuhan Arivazhagan, Vinay Aggarwal, Aaditya~Kumar Singh, and Sunav
  Choudhary.
\newblock Federated learning with personalization layers.
\newblock {\em arXiv preprint arXiv:1912.00818}, 2019.

\bibitem{Assran2019Stochastic}
Mahmoud Assran, Nicolas Loizou, Nicolas Ballas, and Mike Rabbat.
\newblock Stochastic gradient push for distributed deep learning.
\newblock In {\em Int. Conf. on Machine Learning ({ICML})}, volume~97, pages
  344--353, 2019.

\bibitem{ateniese2015hacking}
Giuseppe Ateniese, Luigi~V Mancini, Angelo Spognardi, Antonio Villani, Domenico
  Vitali, and Giovanni Felici.
\newblock Hacking smart machines with smarter ones: How to extract meaningful
  data from machine learning classifiers.
\newblock {\em Int. Journal of Security and Networks}, 10(3):137--150, 2015.

\bibitem{awan2018optimized}
Ammar~Ahmad Awan, Ching-Hsiang Chu, Hari Subramoni, and Dhabaleswar~K Panda.
\newblock Optimized broadcast for deep learning workloads on dense-gpu
  infiniband clusters: Mpi or nccl?
\newblock In {\em European MPI Users' Group Meeting}, pages 1--9, 2018.

\bibitem{bagdasaryan2020backdoor}
Eugene Bagdasaryan, Andreas Veit, Yiqing Hua, Deborah Estrin, and Vitaly
  Shmatikov.
\newblock How to backdoor federated learning.
\newblock In {\em Int. Conf. on Artificial Intelligence and Statistics
  ({AISTATS})}, pages 2938--2948, 2020.

\bibitem{PaddleFL}
Baidu.
\newblock Federated deep learning in paddlepaddle.
\newblock \url{https://github.com/PaddlePaddle/PaddleFL}.
\newblock Online; accessed 16/02/2021.

\bibitem{PaddleInterpretability}
Baidu.
\newblock Paddlepaddle interpretability.
\newblock \url{https://github.com/PaddlePaddle/InterpretDL}.
\newblock Online; accessed 13/03/2021.

\bibitem{beutel2020flower}
Daniel~J Beutel, Taner Topal, Akhil Mathur, Xinchi Qiu, Titouan Parcollet, and
  Nicholas~D Lane.
\newblock Flower: A friendly federated learning research framework.
\newblock {\em arXiv preprint arXiv:2007.14390}, 2020.

\bibitem{bhagoji2019analyzing}
Arjun~Nitin Bhagoji, Supriyo Chakraborty, Prateek Mittal, and Seraphin Calo.
\newblock Analyzing federated learning through an adversarial lens.
\newblock In {\em Int. Conf. on Machine Learning ({ICML})}, pages 634--643,
  2019.

\bibitem{bi2018empirical}
Jingjun Bi and Chongsheng Zhang.
\newblock An empirical comparison on state-of-the-art multi-class imbalance
  learning algorithms and a new diversified ensemble learning scheme.
\newblock {\em Knowledge-Based Systems}, 158:81--93, 2018.

\bibitem{bian2017multi}
Jiang Bian, Haoyi Xiong, Wei Cheng, Wenqing Hu, Zhishan Guo, and Yanjie Fu.
\newblock Multi-party sparse discriminant learning.
\newblock In {\em 2017 IEEE International Conference on Data Mining (ICDM)},
  pages 745--750. IEEE, 2017.

\bibitem{bian2020mp2sda}
Jiang Bian, Haoyi Xiong, Yanjie Fu, Jun Huan, and Zhishan Guo.
\newblock Mp2sda: Multi-party parallelized sparse discriminant learning.
\newblock {\em ACM Transactions on Knowledge Discovery from Data (TKDD)},
  14(3):1--22, 2020.

\bibitem{Bonawitz19}
Keith Bonawitz, Hubert Eichner, Wolfgang Grieskamp, Dzmitry Huba, Alex
  Ingerman, Vladimir Ivanov, Chlo{\'{e}} Kiddon, Jakub Konecn{\'{y}}, Stefano
  Mazzocchi, Brendan McMahan, Timon~Van Overveldt, David Petrou, Daniel Ramage,
  and Jason Roselander.
\newblock Towards federated learning at scale: System design.
\newblock In {\em Machine Learning and Systems ({MLSys})}, 2019.

\bibitem{briggs2020federated}
Christopher Briggs, Zhong Fan, and Peter Andras.
\newblock Federated learning with hierarchical clustering of local updates to
  improve training on non-iid data.
\newblock In {\em Int. Joint Conf. on Neural Networks ({IJCNN})}, pages 1--9.
  IEEE, 2020.

\bibitem{brisimi2018federated}
Theodora~S Brisimi, Ruidi Chen, Theofanie Mela, Alex Olshevsky, Ioannis~Ch
  Paschalidis, and Wei Shi.
\newblock Federated learning of predictive models from federated electronic
  health records.
\newblock {\em Int. journal of Medical Informatics ({IJMI})}, 112:59--67, 2018.

\bibitem{caldas2018leaf}
Sebastian Caldas, Sai Meher~Karthik Duddu, Peter Wu, Tian Li, Jakub
  Kone{\v{c}}n{\`y}, H~Brendan McMahan, Virginia Smith, and Ameet Talwalkar.
\newblock Leaf: A benchmark for federated settings.
\newblock {\em arXiv preprint arXiv:1812.01097}, 2018.

\bibitem{caldas2018expanding}
Sebastian Caldas, Jakub Kone{\v{c}}ny, H~Brendan McMahan, and Ameet Talwalkar.
\newblock Expanding the reach of federated learning by reducing client resource
  requirements.
\newblock {\em arXiv preprint arXiv:1812.07210}, 2018.

\bibitem{canini2012sibyl}
Kevin Canini, Tushar Chandra, Eugene Ie, Jim McFadden, Ken Goldman, Mike
  Gunter, Jeremiah Harmsen, Kristen LeFevre, Dmitry Lepikhin, Tomas~Lloret
  Llinares, et~al.
\newblock Sibyl: A system for large scale supervised machine learning.
\newblock {\em Technical Talk}, 1:113, 2012.

\bibitem{ccatak2015secure}
Ferhat~{\"O}zg{\"u}r {\c{C}}atak.
\newblock Secure multi-party computation based privacy preserving extreme
  learning machine algorithm over vertically distributed data.
\newblock In {\em Int. Conf. on Neural Information Processing ({ICONIP})},
  pages 337--345, 2015.

\bibitem{ccatak2018cpp}
Ferhat~{\"O}zg{\"u}r {\c{C}}atak and Ahmet~Fatih Mustacoglu.
\newblock Cpp-elm: Cryptographically privacy-preserving extreme learning
  machine for cloud systems.
\newblock {\em Int. Journal of Computational Intelligence Systems},
  11(1):33--44, 2018.

\bibitem{chatterjee1977towards}
Samprit Chatterjee and Eugene Seneta.
\newblock Towards consensus: Some convergence theorems on repeated averaging.
\newblock {\em Journal of Applied Probability}, pages 89--97, 1977.

\bibitem{chen2020backdoor}
Chien-Lun Chen, Leana Golubchik, and Marco Paolieri.
\newblock Backdoor attacks on federated meta-learning.
\newblock {\em arXiv preprint arXiv:2006.07026}, 2020.

\bibitem{chen2012diffusion}
Jianshu Chen and Ali~H Sayed.
\newblock Diffusion adaptation strategies for distributed optimization and
  learning over networks.
\newblock {\em IEEE Transactions on Signal Processing}, 60(8):4289--4305, 2012.

\bibitem{Chen2020FedE}
Mingyang Chen, Wen Zhang, Zonggang Yuan, Yantao Jia, and Huajun Chen.
\newblock Fede: Embedding knowledge graphs in federated setting.
\newblock {\em arXiv preprint arXiv:2010.12882}, 2020.

\bibitem{chen2019communication}
Yang Chen, Xiaoyan Sun, and Yaochu Jin.
\newblock Communication-efficient federated deep learning with layerwise
  asynchronous model update and temporally weighted aggregation.
\newblock {\em IEEE Transactions on Neural Networks and Learning Systems},
  31(10):4229--4238, 2019.

\bibitem{chen2020training}
Yu~Chen, Fang Luo, Tong Li, Tao Xiang, Zheli Liu, and Jin Li.
\newblock A training-integrity privacy-preserving federated learning scheme
  with trusted execution environment.
\newblock {\em Information Sciences}, 522:69--79, 2020.

\bibitem{chik2013singapore}
Warren~B Chik.
\newblock The singapore personal data protection act and an assessment of
  future trends in data privacy reform.
\newblock {\em Computer Law \& Security Review}, 29(5):554--575, 2013.

\bibitem{cohen2017emnist}
Gregory Cohen, Saeed Afshar, Jonathan Tapson, and Andre Van~Schaik.
\newblock Emnist: Extending mnist to handwritten letters.
\newblock In {\em Int. Joint Conf. on Neural Networks (IJCNN)}, pages
  2921--2926, 2017.

\bibitem{Uber}
Kate Conger.
\newblock Uber settles data breach investigation for \$148 million.
\newblock
  \url{https://www.nytimes.com/2018/09/26/technology/uber-data-breach.html}.
\newblock Online; accessed 17/02/2021.

\bibitem{UberLoss}
Kate Conger.
\newblock Uber settles data breach investigation for \$148 million, 2018.
\newblock
  \url{https://www.nytimes.com/2018/09/26/technology/uber-data-breach.html}.
\newblock Online; accessed 28/02/2021.

\bibitem{deng2009imagenet}
Jia Deng, Wei Dong, Richard Socher, Li-Jia Li, Kai Li, and Li~Fei-Fei.
\newblock Imagenet: A large-scale hierarchical image database.
\newblock In {\em IEEE conf. on Computer Vision and Pattern Recognition
  ({CVPR})}, pages 248--255, 2009.

\bibitem{deng2020adaptive}
Yuyang Deng, Mohammad~Mahdi Kamani, and Mehrdad Mahdavi.
\newblock Adaptive personalized federated learning.
\newblock {\em arXiv preprint arXiv:2003.13461}, 2020.

\bibitem{dinh2020personalized}
Canh~T Dinh, Nguyen~H Tran, and Tuan~Dung Nguyen.
\newblock Personalized federated learning with moreau envelopes.
\newblock {\em arXiv preprint arXiv:2006.08848}, 2020.

\bibitem{dwork2008differential}
Cynthia Dwork.
\newblock Differential privacy: A survey of results.
\newblock In {\em Int. conf. on theory and applications of models of
  computation}, pages 1--19, 2008.

\bibitem{eddy2004hidden}
Sean~R Eddy.
\newblock What is a hidden markov model?
\newblock {\em Nature biotechnology}, 22(10):1315--1316, 2004.

\bibitem{fang2020local}
Minghong Fang, Xiaoyu Cao, Jinyuan Jia, and Neil Gong.
\newblock Local model poisoning attacks to byzantine-robust federated learning.
\newblock In {\em {USENIX} Security Symposium ({USENIX} Security)}, pages
  1605--1622, 2020.

\bibitem{feng2020multi}
Siwei Feng and Han Yu.
\newblock Multi-participant multi-class vertical federated learning.
\newblock {\em arXiv preprint arXiv:2001.11154}, 2020.

\bibitem{feng2019securegbm}
Zhi Feng, Haoyi Xiong, Chuanyuan Song, Sijia Yang, Baoxin Zhao, Licheng Wang,
  Zeyu Chen, Shengwen Yang, Liping Liu, and Jun Huan.
\newblock Securegbm: Secure multi-party gradient boosting.
\newblock In {\em IEEE Int. Conf. on Big Data ({Big Data})}, pages 1312--1321,
  2019.

\bibitem{fette2011websocket}
Ian Fette and Alexey Melnikov.
\newblock The websocket protocol, 2011.

\bibitem{flynn1972some}
Michael~J Flynn.
\newblock Some computer organizations and their effectiveness.
\newblock {\em IEEE Transactions on Computers}, 100(9):948--960, 1972.

\bibitem{fung2018mitigating}
Clement Fung, Chris~JM Yoon, and Ivan Beschastnikh.
\newblock Mitigating sybils in federated learning poisoning.
\newblock {\em arXiv preprint arXiv:1808.04866}, 2018.

\bibitem{Gaff2014}
B.~M. {Gaff}, H.~E. {Sussman}, and J.~{Geetter}.
\newblock Privacy and big data.
\newblock {\em Computer}, 47(6):7--9, 2014.

\bibitem{ganga2013fault}
K~Ganga and S~Karthik.
\newblock A fault tolerent approach in scientific workflow systems based on
  cloud computing.
\newblock In {\em Int. Conf. on Pattern Recognition, Informatics and Mobile
  Engineering}, pages 387--390, 2013.

\bibitem{geiping2020inverting}
Jonas Geiping, Hartmut Bauermeister, Hannah Dr{\"o}ge, and Michael Moeller.
\newblock Inverting gradients--how easy is it to break privacy in federated
  learning?
\newblock {\em arXiv preprint arXiv:2003.14053}, 2020.

\bibitem{geyer2017differentially}
Robin~C Geyer, Tassilo Klein, and Moin Nabi.
\newblock Differentially private federated learning: A client level
  perspective.
\newblock {\em arXiv preprint arXiv:1712.07557}, 2017.

\bibitem{ring-allreduce}
Andrew Gibiansky.
\newblock Bringing hpc techniques to deep learning.
\newblock
  \url{https://andrew.gibiansky.com/blog/machine-learning/baidu-allreduce/},
  2017.
\newblock Online; accessed 2020-08-12.

\bibitem{gilpin2018explaining}
Leilani~H Gilpin, David Bau, Ben~Z Yuan, Ayesha Bajwa, Michael Specter, and
  Lalana Kagal.
\newblock Explaining explanations: An overview of interpretability of machine
  learning.
\newblock In {\em IEEE Int. Conf. on Data Science and Advanced Analytics
  ({DSAA})}, pages 80--89. IEEE, 2018.

\bibitem{goodfellow2016deep}
Ian Goodfellow, Yoshua Bengio, Aaron Courville, and Yoshua Bengio.
\newblock {\em Deep learning}, volume~1.
\newblock MIT press Cambridge, 2016.

\bibitem{Goodfellow2015Explaining}
Ian Goodfellow, Jonathon Shlens, and Christian Szegedy.
\newblock Explaining and harnessing adversarial examples.
\newblock In {\em Int. Conf. on Learning Representations ({ICLR})}, 2015.

\bibitem{TFF}
Google.
\newblock Tensorflow federated: Machine learning on decentralized data.
\newblock \url{https://www.tensorflow.org/federated}.
\newblock Online; accessed 16/02/2021.

\bibitem{gropp1999using}
William Gropp, William~D Gropp, Ewing Lusk, Anthony Skjellum, and Argonne
  Distinguished Fellow Emeritus~Ewing Lusk.
\newblock {\em Using MPI: portable parallel programming with the
  message-passing interface}, volume~1.
\newblock MIT press, 1999.

\bibitem{haddadpour2020federated}
Farzin Haddadpour, Mohammad~Mahdi Kamani, Aryan Mokhtari, and Mehrdad Mahdavi.
\newblock Federated learning with compression: Unified analysis and sharp
  guarantees.
\newblock {\em arXiv preprint arXiv:2007.01154}, 2020.

\bibitem{hao2019towards}
Meng Hao, Hongwei Li, Guowen Xu, Sen Liu, and Haomiao Yang.
\newblock Towards efficient and privacy-preserving federated deep learning.
\newblock In {\em IEEE Int. Conf. on Communications ({ICC})}, pages 1--6, 2019.

\bibitem{hardy2017private}
Stephen Hardy, Wilko Henecka, Hamish Ivey-Law, Richard Nock, Giorgio Patrini,
  Guillaume Smith, and Brian Thorne.
\newblock Private federated learning on vertically partitioned data via entity
  resolution and additively homomorphic encryption.
\newblock {\em arXiv preprint arXiv:1711.10677}, 2017.

\bibitem{He2020Group}
Chaoyang He, Murali Annavaram, and Salman Avestimehr.
\newblock Group knowledge transfer: Collaborative training of large cnns on the
  edge.
\newblock {\em arXiv preprint arXiv:2007.14513}, 2020.

\bibitem{he2020towards}
Chaoyang He, Murali Annavaram, and Salman Avestimehr.
\newblock Towards non-iid and invisible data with fednas: Federated deep
  learning via neural architecture search.
\newblock {\em arXiv preprint arXiv:2004.08546}, 2020.

\bibitem{he2021fedgraphnn}
Chaoyang He, Keshav Balasubramanian, Emir Ceyani, Yu~Rong, Peilin Zhao, Junzhou
  Huang, Murali Annavaram, and Salman Avestimehr.
\newblock Fedgraphnn: A federated learning system and benchmark for graph
  neural networks.
\newblock {\em arXiv preprint arXiv:2104.07145}, 2021.

\bibitem{He2021SpreadGNN}
Chaoyang He, Emir Ceyani, Keshav Balasubramanian, Murali Annavaram, and Salman
  Avestimehr.
\newblock Spreadgnn: Serverless multi-task federated learning for graph neural
  networks.
\newblock {\em arXiv preprint arXiv:2106.02743}, 2021.

\bibitem{He2021PipeTransformer}
Chaoyang He, Shen Li, Mahdi Soltanolkotabi, and Salman Avestimehr.
\newblock Pipetransformer: Automated elastic pipelining for distributed
  training of large-scale models.
\newblock In {\em Int. Conf. on Machine Learning}, volume 139 of {\em Machine
  Learning Research}, pages 4150--4159, 2021.

\bibitem{he2020fedml}
Chaoyang He, Songze Li, Jinhyun So, Xiao Zeng, Mi~Zhang, Hongyi Wang, Xiaoyang
  Wang, Praneeth Vepakomma, Abhishek Singh, Hang Qiu, et~al.
\newblock Fedml: A research library and benchmark for federated machine
  learning.
\newblock {\em arXiv preprint arXiv:2007.13518}, 2020.

\bibitem{hefedcv}
Chaoyang He, Alay~Dilipbhai Shah, Zhenheng Tang, Di~Fan, Adarshan~Naiynar
  Sivashunmugam, Keerti Bhogaraju, Mita Shimpi, Li~Shen, Xiaowen Chu, Mahdi
  Soltanolkotabi, et~al.
\newblock Fedcv: A federated learning framework for diverse computer vision
  tasks.

\bibitem{He2019Central}
Chaoyang He, Conghui Tan, Hanlin Tang, Shuang Qiu, and Ji~Liu.
\newblock Central server free federated learning over single-sided trust social
  networks.
\newblock {\em arXiv preprint arXiv:1910.04956}, 2019.

\bibitem{He2020CVPR}
Chaoyang He, Haishan Ye, Li~Shen, and Tong Zhang.
\newblock Milenas: Efficient neural architecture search via mixed-level
  reformulation.
\newblock In {\em IEEE/CVF Conf. on Computer Vision and Pattern Recognition
  ({CVPR})}, 2020.

\bibitem{He2009Imbalanced}
Haibo He and Edwardo~A. Garcia.
\newblock Learning from imbalanced data.
\newblock {\em IEEE Transactions on Knowledge and Data Engineering ({TKDE})},
  21(9):1263--1284, 2009.

\bibitem{he2020secure}
Lie He, Sai~Praneeth Karimireddy, and Martin Jaggi.
\newblock Secure byzantine-robust machine learning.
\newblock {\em arXiv preprint arXiv:2006.04747}, 2020.

\bibitem{hitaj2017deep}
Briland Hitaj, Giuseppe Ateniese, and Fernando Perez-Cruz.
\newblock Deep models under the gan: information leakage from collaborative
  deep learning.
\newblock In {\em ACM SIGSAC Conference on Computer and Communications
  Security}, pages 603--618, 2017.

\bibitem{Hu2019Decentralized}
Chenghao Hu, Jingyan Jiang, and Zhi Wang.
\newblock Decentralized federated learning: {A} segmented gossip approach.
\newblock {\em arXiv preprint arXiv:1908.07782}, 2019.

\bibitem{hu2020fedmgda}
Zeou Hu, Kiarash Shaloudegi, Guojun Zhang, and Yaoliang Yu.
\newblock Fedmgda+: Federated learning meets multi-objective optimization.
\newblock {\em arXiv preprint arXiv:2006.11489}, 2020.

\bibitem{huang2018gpipe}
Yanping Huang, Youlong Cheng, Ankur Bapna, Orhan Firat, Mia~Xu Chen, Dehao
  Chen, HyoukJoong Lee, Jiquan Ngiam, Quoc~V Le, Yonghui Wu, et~al.
\newblock Gpipe: Efficient training of giant neural networks using pipeline
  parallelism.
\newblock {\em arXiv preprint arXiv:1811.06965}, 2018.

\bibitem{ivkin2019communication}
Nikita Ivkin, Daniel Rothchild, Enayat Ullah, Vladimir Braverman, Ion Stoica,
  and Raman Arora.
\newblock Communication-efficient distributed sgd with sketching.
\newblock {\em arXiv preprint arXiv:1903.04488}, 2019.

\bibitem{jiang2018sketchml}
Jiawei Jiang, Fangcheng Fu, Tong Yang, and Bin Cui.
\newblock Sketchml: Accelerating distributed machine learning with data
  sketches.
\newblock In {\em Int. Conf. on Management of Data}, pages 1269--1284, 2018.

\bibitem{jiang2020decentralized}
Jing Jiang, Shaoxiong Ji, and Guodong Long.
\newblock Decentralized knowledge acquisition for mobile internet applications.
\newblock {\em World Wide Web}, pages 1--17, 2020.

\bibitem{jiang2020federated}
Meng Jiang, Taeho Jung, Ryan Karl, and Tong Zhao.
\newblock Federated dynamic gnn with secure aggregation.
\newblock {\em arXiv preprint arXiv:2009.07351}, 2020.

\bibitem{kairouz2019advances}
Peter Kairouz, H~Brendan McMahan, Brendan Avent, Aur{\'e}lien Bellet, Mehdi
  Bennis, Arjun~Nitin Bhagoji, Keith Bonawitz, Zachary Charles, Graham Cormode,
  Rachel Cummings, et~al.
\newblock Advances and open problems in federated learning.
\newblock {\em arXiv preprint arXiv:1912.04977}, 2019.

\bibitem{mcmahan2021advances}
Peter Kairouz, H.~Brendan McMahan, Aurélien~Bellet Brendan~Avent, Arjun
  Nitin~Bhagoji Mehdi~Bennis, Keith Bonawitz, Zachary Charles, Graham Cormode,
  Rachel Cummings, Rafael~G.L. D'Oliveira, Salim~El Rouayheb, David Evans, Josh
  Gardner, Zachary Garrett, Adrià Gascón, Phillip B.~Gibbons Badih~Ghazi,
  Marco Gruteser, Zaid Harchaoui, Chaoyang He, Lie He, Zhouyuan Huo, Ben
  Hutchinson, Justin Hsu, Martin Jaggi, Tara Javidi, Gauri Joshi, Mikhail
  Khodak, Jakub Konečný, Aleksandra Korolova, Farinaz Koushanfar, Sanmi
  Koyejo, Tancrède Lepoint, Yang Liu, Prateek Mittal, Mehryar Mohri, Richard
  Nock, Ayfer Özgür, Rasmus Pagh, Mariana Raykova, Hang Qi, Daniel Ramage,
  Ramesh Raskar, Dawn Song, Weikang Song, Sebastian~U. Stich, Ziteng Sun,
  Ananda~Theertha Suresh, Florian Tramèr, Praneeth Vepakomma, Jianyu Wang,
  Li~Xiong, Zheng Xu, Qiang Yang, Felix~X. Yu, Han Yu, and Sen Zhao.
\newblock Advances and open problems in federated learning.
\newblock {\em Foundations and Trends{\textregistered} in Machine Learning},
  14(1), 2021.

\bibitem{karimireddy2020scaffold}
Sai~Praneeth Karimireddy, Satyen Kale, Mehryar Mohri, Sashank Reddi, Sebastian
  Stich, and Ananda~Theertha Suresh.
\newblock Scaffold: Stochastic controlled averaging for federated learning.
\newblock In {\em Int. Conf. on Machine Learning ({ICML})}, pages 5132--5143,
  2020.

\bibitem{karimireddy2019error}
Sai~Praneeth Karimireddy, Quentin Rebjock, Sebastian Stich, and Martin Jaggi.
\newblock Error feedback fixes signsgd and other gradient compression schemes.
\newblock In {\em Int. Conf. on Machine Learning ({ICML})}, pages 3252--3261,
  2019.

\bibitem{katevas2020policy}
Kleomenis Katevas, Eugene Bagdasaryan, Jason Waterman, Mohamad Mounir~Safadieh,
  Eleanor Birrell, Hamed Haddadi, and Deborah Estrin.
\newblock Policy-based federated learning.
\newblock {\em arXiv e-prints}, pages arXiv--2003, 2020.

\bibitem{Ke2021Federated}
Chuyang Ke and Jean Honorio.
\newblock Federated myopic community detection with one-shot communication.
\newblock {\em arXiv preprint arXiv:2106.07255}, 2021.

\bibitem{Kholod2021}
Ivan Kholod, Evgeny Yanaki, Dmitry Fomichev, Evgeniy Shalugin, Evgenia
  Novikova, Evgeny Filippov, and Mats Nordlund.
\newblock Open-source federated learning frameworks for iot: {A} comparative
  review and analysis.
\newblock {\em Sensors}, 21(1):167, 2021.

\bibitem{konevcny2016federated}
Jakub Kone{\v{c}}n{\`y}, H~Brendan McMahan, Felix~X Yu, Peter Richt{\'a}rik,
  Ananda~Theertha Suresh, and Dave Bacon.
\newblock Federated learning: Strategies for improving communication
  efficiency.
\newblock {\em arXiv preprint arXiv:1610.05492}, 2016.

\bibitem{kulkarni2020survey}
Viraj Kulkarni, Milind Kulkarni, and Aniruddha Pant.
\newblock Survey of personalization techniques for federated learning.
\newblock In {\em World Conf. on Smart Trends in Systems, Security and
  Sustainability ({WorldS4})}, pages 794--797, 2020.

\bibitem{lalitha2019peer}
Anusha Lalitha, Osman~Cihan Kilinc, Tara Javidi, and Farinaz Koushanfar.
\newblock Peer-to-peer federated learning on graphs.
\newblock {\em arXiv preprint arXiv:1901.11173}, 2019.

\bibitem{li2019survey}
Qinbin Li, Zeyi Wen, Zhaomin Wu, Sixu Hu, Naibo Wang, and Bingsheng He.
\newblock A survey on federated learning systems: vision, hype and reality for
  data privacy and protection.
\newblock {\em arXiv preprint arXiv:1907.09693}, 2019.

\bibitem{li2019abnormal}
Suyi Li, Yong Cheng, Yang Liu, Wei Wang, and Tianjian Chen.
\newblock Abnormal client behavior detection in federated learning.
\newblock {\em arXiv preprint arXiv:1910.09933}, 2019.

\bibitem{li2020federated}
Tian Li, Anit~Kumar Sahu, Ameet Talwalkar, and Virginia Smith.
\newblock Federated learning: Challenges, methods, and future directions.
\newblock {\em IEEE Signal Processing Magazine}, 37(3):50--60, 2020.

\bibitem{Li2020}
Tian Li, Anit~Kumar Sahu, Manzil Zaheer, Maziar Sanjabi, Ameet Talwalkar, and
  Virginia Smith.
\newblock Federated optimization in heterogeneous networks.
\newblock In {\em Machine Learning and Systems}, volume~2, pages 429--450,
  2020.

\bibitem{li2019fair}
Tian Li, Maziar Sanjabi, Ahmad Beirami, and Virginia Smith.
\newblock Fair resource allocation in federated learning.
\newblock {\em arXiv preprint arXiv:1905.10497}, 2019.

\bibitem{li2020backdoor}
Yiming Li, Baoyuan Wu, Yong Jiang, Zhifeng Li, and Shu-Tao Xia.
\newblock Backdoor learning: A survey.
\newblock {\em arXiv preprint arXiv:2007.08745}, 2020.

\bibitem{li2019quantification}
Zhaorui Li, Zhicong Huang, Chaochao Chen, and Cheng Hong.
\newblock Quantification of the leakage in federated learning.
\newblock {\em arXiv preprint arXiv:1910.05467}, 2019.

\bibitem{lian2017can}
Xiangru Lian, Ce~Zhang, Huan Zhang, Cho-Jui Hsieh, Wei Zhang, and Ji~Liu.
\newblock Can decentralized algorithms outperform centralized algorithms? a
  case study for decentralized parallel stochastic gradient descent.
\newblock In {\em Advances in Neural Information Processing Systems
  ({NeurIPS})}, pages 5330--5340, 2017.

\bibitem{liang2020exploring}
Zhicong Liang, Bao Wang, Quanquan Gu, Stanley Osher, and Yuan Yao.
\newblock Exploring private federated learning with laplacian smoothing.
\newblock {\em arXiv preprint arXiv:2005.00218}, 2020.

\bibitem{liaqat2017federated}
Misbah Liaqat, Victor Chang, Abdullah Gani, Siti~Hafizah Ab~Hamid, Muhammad
  Toseef, Umar Shoaib, and Rana~Liaqat Ali.
\newblock Federated cloud resource management: Review and discussion.
\newblock {\em Journal of Network and Computer Applications}, 77:87--105, 2017.

\bibitem{lim2020federated}
Wei Yang~Bryan Lim, Nguyen~Cong Luong, Dinh~Thai Hoang, Yutao Jiao, Ying-Chang
  Liang, Qiang Yang, Dusit Niyato, and Chunyan Miao.
\newblock Federated learning in mobile edge networks: A comprehensive survey.
\newblock {\em IEEE Communications Surveys \& Tutorials}, 22(3):2031--2063,
  2020.

\bibitem{Lin2021FedNLP}
Bill~Yuchen Lin, Chaoyang He, Zihang Zeng, Hulin Wang, Yufen Huang, Mahdi
  Soltanolkotabi, Xiang Ren, and Salman Avestimehr.
\newblock Fednlp: {A} research platform for federated learning in natural
  language processing.
\newblock {\em arXiv preprint arXiv:2104.08815}, 2021.

\bibitem{lin2019free}
Jierui Lin, Min Du, and Jian Liu.
\newblock Free-riders in federated learning: Attacks and defenses.
\newblock {\em arXiv preprint arXiv:1911.12560}, 2019.

\bibitem{lin2020improving}
Yilun Lin, Chaochao Chen, Cen Chen, and Li~Wang.
\newblock Improving federated relational data modeling via basis alignment and
  weight penalty.
\newblock {\em arXiv preprint arXiv:2011.11369}, 2020.

\bibitem{liu2020two}
Ji~Liu, Carlyna Bondiombouy, Lei Mo, and Patrick Valduriez.
\newblock Two-phase scheduling for efficient vehicle sharing.
\newblock {\em IEEE Transactions on Intelligent Transportation Systems
  ({TITS})}, 2020.

\bibitem{liu2016multi}
Ji~Liu, Esther Pacitti, Patrick Valduriez, Daniel De~Oliveira, and Marta
  Mattoso.
\newblock Multi-objective scheduling of scientific workflows in multisite
  clouds.
\newblock {\em Future Generation Computer Systems}, 63:76--95, 2016.

\bibitem{liu2015survey}
Ji~Liu, Esther Pacitti, Patrick Valduriez, and Marta Mattoso.
\newblock A survey of data-intensive scientific workflow management.
\newblock {\em Journal of Grid Computing}, 13(4):457--493, 2015.

\bibitem{liu2018efficient}
Ji~Liu, Luis Pineda, Esther Pacitti, Alexandru Costan, Patrick Valduriez,
  Gabriel Antoniu, and Marta Mattoso.
\newblock Efficient scheduling of scientific workflows using hot metadata in a
  multisite cloud.
\newblock {\em IEEE Transactions on Knowledge and Data Engineering ({TKDE})},
  31(10):1940--1953, 2018.

\bibitem{liu2020client}
Lumin Liu, Jun Zhang, SH~Song, and Khaled~B Letaief.
\newblock Client-edge-cloud hierarchical federated learning.
\newblock In {\em IEEE Int. Conf. on Communications ({ICC})}, pages 1--6, 2020.

\bibitem{liu2020fedsel}
Ruixuan Liu, Yang Cao, Masatoshi Yoshikawa, and Hong Chen.
\newblock Fedsel: Federated sgd under local differential privacy with top-k
  dimension selection.
\newblock In {\em Int. Conf. on Database Systems for Advanced Applications},
  pages 485--501, 2020.

\bibitem{liu2020fedvision}
Yang Liu, Anbu Huang, Yun Luo, He~Huang, Youzhi Liu, Yuanyuan Chen, Lican Feng,
  Tianjian Chen, Han Yu, and Qiang Yang.
\newblock Fedvision: An online visual object detection platform powered by
  federated learning.
\newblock In {\em AAAI Conf. on Artificial Intelligence}, volume~34, pages
  13172--13179, 2020.

\bibitem{liu2019communication}
Yang Liu, Yan Kang, Xinwei Zhang, Liping Li, Yong Cheng, Tianjian Chen, Mingyi
  Hong, and Qiang Yang.
\newblock A communication efficient collaborative learning framework for
  distributed features.
\newblock {\em arXiv preprint arXiv:1912.11187}, 2019.

\bibitem{lo2021architectural}
Sin~Kit Lo, Qinghua Lu, Liming Zhu, Hye-young Paik, Xiwei Xu, and Chen Wang.
\newblock Architectural patterns for the design of federated learning systems.
\newblock {\em arXiv preprint arXiv:2101.02373}, 2021.

\bibitem{luo2020hfel}
Siqi Luo, Xu~Chen, Qiong Wu, Zhi Zhou, and Shuai Yu.
\newblock Hfel: Joint edge association and resource allocation for
  cost-efficient hierarchical federated edge learning.
\newblock {\em IEEE Transactions on Wireless Communications},
  19(10):6535--6548, 2020.

\bibitem{luo2020exploiting}
Xinjian Luo and Xiangqi Zhu.
\newblock Exploiting defenses against gan-based feature inference attacks in
  federated learning.
\newblock {\em arXiv preprint arXiv:2004.12571}, 2020.

\bibitem{lyu2020threats}
Lingjuan Lyu, Han Yu, and Qiang Yang.
\newblock Threats to federated learning: A survey.
\newblock {\em arXiv preprint arXiv:2003.02133}, 2020.

\bibitem{lyu2020towards}
Lingjuan Lyu, Jiangshan Yu, Karthik Nandakumar, Yitong Li, Xingjun Ma, Jiong
  Jin, Han Yu, and Kee~Siong Ng.
\newblock Towards fair and privacy-preserving federated deep models.
\newblock {\em IEEE Transactions on Parallel and Distributed Systems ({TPDS})},
  31(11):2524--2541, 2020.

\bibitem{Ma2019}
Yanjun Ma, Dianhai~Yu adn Tian~Wu, and Haifeng Wang.
\newblock Paddlepaddle: An open-source deep learning platform from industrial
  practice.
\newblock {\em Frontiers of Data and Computing}, 1(1):105, 2019.

\bibitem{Malekijoo2021FEDZIP}
Amirhossein Malekijoo, Mohammad~Javad Fadaeieslam, Hanieh Malekijou, Morteza
  Homayounfar, Farshid Alizadeh{-}Shabdiz, and Reza Rawassizadeh.
\newblock {FEDZIP:} {A} compression framework for communication-efficient
  federated learning.
\newblock {\em arXiv preprint arXiv:2102.01593}, 2021.

\bibitem{mandal2019privfl}
Kalikinkar Mandal and Guang Gong.
\newblock {PrivFL}: Practical privacy-preserving federated regressions on
  high-dimensional data over mobile networks.
\newblock In {\em ACM SIGSAC Conf. on Cloud Computing Security Workshop}, pages
  57--68, 2019.

\bibitem{mckeen2013innovative}
Frank McKeen, Ilya Alexandrovich, Alex Berenzon, Carlos~V. Rozas, Hisham Shafi,
  Vedvyas Shanbhogue, and Uday~R. Savagaonkar.
\newblock Innovative instructions and software model for isolated execution.
\newblock In {\em Int. Workshop on Hardware and Architectural Support for
  Security and Privacy}, 2013.

\bibitem{mcmahan2017communication}
Brendan McMahan, Eider Moore, Daniel Ramage, Seth Hampson, and Blaise~Aguera
  y~Arcas.
\newblock Communication-efficient learning of deep networks from decentralized
  data.
\newblock In {\em Int. Conf. on Artificial Intelligence and Statistics
  ({AISTATS})}, pages 1273--1282, 2017.

\bibitem{mcmahan2017learning}
H~Brendan McMahan, Daniel Ramage, Kunal Talwar, and Li~Zhang.
\newblock Learning differentially private recurrent language models.
\newblock {\em arXiv preprint arXiv:1710.06963}, 2017.

\bibitem{McMahan2018}
H.~Brendan McMahan, Daniel Ramage, Kunal Talwar, and Li~Zhang.
\newblock Learning differentially private recurrent language models.
\newblock In {\em Int. Conf. on Learning Representations ({ICLR})}, 2018.

\bibitem{mei2019sgnn}
Guangxu Mei, Ziyu Guo, Shijun Liu, and Li~Pan.
\newblock Sgnn: A graph neural network based federated learning approach by
  hiding structure.
\newblock In {\em IEEE Int. Conf. on Big Data (Big Data)}, pages 2560--2568,
  2019.

\bibitem{melis2019exploiting}
Luca Melis, Congzheng Song, Emiliano De~Cristofaro, and Vitaly Shmatikov.
\newblock Exploiting unintended feature leakage in collaborative learning.
\newblock In {\em IEEE Symposium on Security and Privacy ({SP})}, pages
  691--706, 2019.

\bibitem{Meng2021Cross}
Chuizheng Meng, Sirisha Rambhatla, and Yan Liu.
\newblock Cross-node federated graph neural network for spatio-temporal data
  modeling.
\newblock In {\em ACM SIGKDD Conference on Knowledge Discovery and Data Mining
  ({KDD})}, 2021.
\newblock To appear.

\bibitem{mhaisen2021optimal}
Naram Mhaisen, Alaa Awad, Amr Mohamed, Aiman Erbad, and Mohsen Guizani.
\newblock Optimal user-edge assignment in hierarchical federated learning based
  on statistical properties and network topology constraints.
\newblock {\em IEEE Transactions on Network Science and Engineering}, 2021.

\bibitem{mo2019efficient}
Fan Mo and Hamed Haddadi.
\newblock Efficient and private federated learning using tee.
\newblock In {\em EuroSys}, 2019.

\bibitem{mohri2019agnostic}
Mehryar Mohri, Gary Sivek, and Ananda~Theertha Suresh.
\newblock Agnostic federated learning.
\newblock In {\em Int. Conf. on Machine Learning ({ICML})}, pages 4615--4625,
  2019.

\bibitem{mothukuri2021survey}
Viraaji Mothukuri, Reza~M Parizi, Seyedamin Pouriyeh, Yan Huang, Ali
  Dehghantanha, and Gautam Srivastava.
\newblock A survey on security and privacy of federated learning.
\newblock {\em Future Generation Computer Systems}, 115:619--640, 2021.

\bibitem{munoz2019byzantine}
Luis Mu{\~n}oz-Gonz{\'a}lez, Kenneth~T Co, and Emil~C Lupu.
\newblock Byzantine-robust federated machine learning through adaptive model
  averaging.
\newblock {\em arXiv preprint arXiv:1909.05125}, 2019.

\bibitem{narayanan2019pipedream}
Deepak Narayanan, Aaron Harlap, Amar Phanishayee, Vivek Seshadri, Nikhil~R
  Devanur, Gregory~R Ganger, Phillip~B Gibbons, and Matei Zaharia.
\newblock Pipedream: generalized pipeline parallelism for dnn training.
\newblock In {\em ACM Symposium on Operating Systems Principles}, pages 1--15,
  2019.

\bibitem{ochiai2019real}
Keiichi Ochiai, Kohei Senkawa, Naoki Yamamoto, Yuya Tanaka, and Yusuke
  Fukazawa.
\newblock Real-time on-device troubleshooting recommendation for smartphones.
\newblock In {\em ACM SIGKDD Int. Conf. on Knowledge Discovery \& Data Mining},
  pages 2783--2791, 2019.

\bibitem{GDPR}
{Official Journal of the European Union}.
\newblock General data protection regulation.
\newblock
  \url{https://eur-lex.europa.eu/legal-content/EN/TXT/PDF/?uri=CELEX:32016R0679}.
\newblock Online; accessed 12/02/2021.

\bibitem{ohrimenko2016oblivious}
Olga Ohrimenko, Felix Schuster, C{\'e}dric Fournet, Aastha Mehta, Sebastian
  Nowozin, Kapil Vaswani, and Manuel Costa.
\newblock Oblivious multi-party machine learning on trusted processors.
\newblock In {\em $\{$USENIX$\}$ Security Symposium ($\{$USENIX$\}$ Security)},
  pages 619--636, 2016.

\bibitem{oksuz2020imbalance}
Kemal Oksuz, Baris~Can Cam, Sinan Kalkan, and Emre Akbas.
\newblock Imbalance problems in object detection: A review.
\newblock {\em IEEE transactions on pattern analysis and machine intelligence},
  2020.

\bibitem{Pysyft}
OpenMined.
\newblock Pysyft.
\newblock \url{https://github.com/OpenMined/PySyft}.
\newblock Online; accessed 22/02/2021.

\bibitem{PaddleHub}
{PaddlePaddle, Baidu}.
\newblock Paddlehub.
\newblock \url{https://github.com/PaddlePaddle/PaddleHub}.
\newblock Online; accessed 01/10/2021.

\bibitem{paillier1999public}
Pascal Paillier.
\newblock Public-key cryptosystems based on composite degree residuosity
  classes.
\newblock In {\em Int. Conf. on the theory and applications of cryptographic
  techniques}, pages 223--238, 1999.

\bibitem{pan2009survey}
Sinno~Jialin Pan and Qiang Yang.
\newblock A survey on transfer learning.
\newblock {\em IEEE Transactions on Knowledge and Data Engineering ({TKDE})},
  22(10):1345--1359, 2009.

\bibitem{peng2021federated}
Hao Peng, Haoran Li, Yangqiu Song, Vincent Zheng, and Jianxin Li.
\newblock Federated knowledge graphs embedding.
\newblock In {\em {ACM} Int. Conf. on Information and Knowledge Management
  ({CIKM})}, pages 1--10, 2021.

\bibitem{phan2020scalable}
Hai Phan, My~T Thai, Han Hu, Ruoming Jin, Tong Sun, and Dejing Dou.
\newblock Scalable differential privacy with certified robustness in
  adversarial learning.
\newblock In {\em Int. Conf. on Machine Learning ({ICML})}, pages 7683--7694,
  2020.

\bibitem{pillutla2019robust}
Krishna Pillutla, Sham~M Kakade, and Zaid Harchaoui.
\newblock Robust aggregation for federated learning.
\newblock {\em arXiv preprint arXiv:1912.13445}, 2019.

\bibitem{pineda2016managing}
Luis Pineda-Morales, Ji~Liu, Alexandru Costan, Esther Pacitti, Gabriel Antoniu,
  Patrick Valduriez, and Marta Mattoso.
\newblock Managing hot metadata for scientific workflows on multisite clouds.
\newblock In {\em IEEE Int. Conf. on Big Data ({Big Data})}, pages 390--397,
  2016.

\bibitem{Pytorch}
Pytorch.
\newblock Pytorch.
\newblock \url{https://pytorch.org/}.
\newblock Online; accessed 13/03/2021.

\bibitem{robbins1951stochastic}
Herbert Robbins and Sutton Monro.
\newblock A stochastic approximation method.
\newblock {\em The annals of mathematical statistics}, pages 400--407, 1951.

\bibitem{Romanini2021}
Daniele Romanini, Adam~James Hall, Pavlos Papadopoulos, Tom Titcombe, Abbas
  Ismail, Tudor Cebere, Robert Sandmann, Robin Roehm, and Michael~A. Hoeh.
\newblock Pyvertical: {A} vertical federated learning framework for
  multi-headed splitnn.
\newblock {\em arXiv preprint arXiv:2104.00489}, 2021.

\bibitem{rothchild2020fetchsgd}
Daniel Rothchild, Ashwinee Panda, Enayat Ullah, Nikita Ivkin, Ion Stoica,
  Vladimir Braverman, Joseph Gonzalez, and Raman Arora.
\newblock Fetchsgd: Communication-efficient federated learning with sketching.
\newblock In {\em Int. Conf. on Machine Learning ({ICML})}, pages 8253--8265,
  2020.

\bibitem{ryffel2018generic}
Theo Ryffel, Andrew Trask, Morten Dahl, Bobby Wagner, Jason Mancuso, Daniel
  Rueckert, and Jonathan Passerat-Palmbach.
\newblock A generic framework for privacy preserving deep learning.
\newblock {\em arXiv preprint arXiv:1811.04017}, 2018.

\bibitem{sabater2020distributed}
C{\'e}sar Sabater, Aur{\'e}lien Bellet, and Jan Ramon.
\newblock Distributed differentially private averaging with improved utility
  and robustness to malicious parties.
\newblock {\em arXiv preprint arXiv:2006.07218}, 2020.

\bibitem{GoogleLoss}
Adam Satariano.
\newblock Google is fined \$57 million under europe’s data privacy law.
\newblock
  \url{https://www.nytimes.com/2019/01/21/technology/google-europe-gdpr-fine.html}.
\newblock Online; accessed 28/02/2021.

\bibitem{sayed2014adaptation}
Ali~H Sayed.
\newblock Adaptation, learning, and optimization over networks.
\newblock {\em Foundations and Trends in Machine Learning},
  7(ARTICLE):311--801, 2014.

\bibitem{sayed2013diffusion}
Ali~H Sayed, Sheng-Yuan Tu, Jianshu Chen, Xiaochuan Zhao, and Zaid~J Towfic.
\newblock Diffusion strategies for adaptation and learning over networks: an
  examination of distributed strategies and network behavior.
\newblock {\em IEEE Signal Processing Magazine}, 30(3):155--171, 2013.

\bibitem{seif2020wireless}
Mohamed Seif, Ravi Tandon, and Ming Li.
\newblock Wireless federated learning with local differential privacy.
\newblock In {\em IEEE Int. Symposium on Information Theory (ISIT)}, pages
  2604--2609, 2020.

\bibitem{seneta2006non}
Eugene Seneta.
\newblock {\em Non-negative matrices and Markov chains}.
\newblock Springer Science \& Business Media, 2006.

\bibitem{shakespeare2007complete}
William Shakespeare.
\newblock {\em The complete works of William Shakespeare}.
\newblock Wordsworth Editions, 2007.

\bibitem{shlezinger2020federated}
Nir Shlezinger, Mingzhe Chen, Yonina~C Eldar, H~Vincent Poor, and Shuguang Cui.
\newblock Federated learning with quantization constraints.
\newblock In {\em IEEE Int. Conf. on Acoustics, Speech and Signal Processing
  ({ICASSP})}, pages 8851--8855, 2020.

\bibitem{shlezinger2020uveqfed}
Nir Shlezinger, Mingzhe Chen, Yonina~C Eldar, H~Vincent Poor, and Shuguang Cui.
\newblock Uveqfed: Universal vector quantization for federated learning.
\newblock {\em IEEE Transactions on Signal Processing}, 2020.

\bibitem{FacebookLoss}
JASON SILVERSTEIN.
\newblock Hundreds of millions of facebook user records were exposed on amazon
  cloud server.
\newblock
  \url{https://www.cbsnews.com/news/millions-facebook-user-records-exposed-amazon-cloud-server/}.
\newblock Online; accessed 28/02/2021.

\bibitem{spring2019compressing}
Ryan Spring, Anastasios Kyrillidis, Vijai Mohan, and Anshumali Shrivastava.
\newblock Compressing gradient optimizers via count-sketches.
\newblock In {\em Int. Conf. on Machine Learning ({ICML})}, pages 5946--5955,
  2019.

\bibitem{CCL}
{Standing Committee of the National People's Congress}.
\newblock Cybersecurity law of the people’s republic of china.
\newblock
  \url{https://www.newamerica.org/cybersecurity-initiative/digichina/blog/translation-cybersecurity-law-peoples-republic-china/}.
\newblock Online; accessed 22/02/2021.

\bibitem{Stich2018}
Sebastian~U Stich, Jean-Baptiste Cordonnier, and Martin Jaggi.
\newblock Sparsified {SGD} with memory.
\newblock In {\em Advances in Neural Information Processing Systems
  ({NeurIPS})}, volume~31, 2018.

\bibitem{sun2020adaptive}
Haijian Sun, Xiang Ma, and Rose~Qingyang Hu.
\newblock Adaptive federated learning with gradient compression in uplink noma.
\newblock {\em IEEE Transactions on Vehicular Technology}, 2020.

\bibitem{sun2019can}
Ziteng Sun, Peter Kairouz, Ananda~Theertha Suresh, and H~Brendan McMahan.
\newblock Can you really backdoor federated learning?
\newblock {\em arXiv preprint arXiv:1911.07963}, 2019.

\bibitem{Suzumura2019Towards}
Toyotaro Suzumura, Yi~Zhou, Nathalie Barcardo, Guangnan Ye, Keith Houck, Ryo
  Kawahara, Ali Anwar, Lucia~Larise Stavarache, Daniel Klyashtorny, Heiko
  Ludwig, and Kumar Bhaskaran.
\newblock Towards federated graph learning for collaborative financial crimes
  detection.
\newblock {\em arXiv preprint arXiv:1909.12946}, 2019.

\bibitem{tolpegin2020data}
Vale Tolpegin, Stacey Truex, Mehmet~Emre Gursoy, and Ling Liu.
\newblock Data poisoning attacks against federated learning systems.
\newblock In {\em European Symposium on Research in Computer Security}, pages
  480--501. Springer, 2020.

\bibitem{triastcyn2020federated}
Aleksei Triastcyn and Boi Faltings.
\newblock Federated generative privacy.
\newblock {\em IEEE Intelligent Systems}, 35(4):50--57, 2020.

\bibitem{truex2019hybrid}
Stacey Truex, Nathalie Baracaldo, Ali Anwar, Thomas Steinke, Heiko Ludwig, Rui
  Zhang, and Yi~Zhou.
\newblock A hybrid approach to privacy-preserving federated learning.
\newblock In {\em ACM Workshop on Artificial Intelligence and Security}, pages
  1--11, 2019.

\bibitem{vanhaesebrouck2017decentralized}
Paul Vanhaesebrouck, Aur{\'e}lien Bellet, and Marc Tommasi.
\newblock Decentralized collaborative learning of personalized models over
  networks.
\newblock In {\em Int. Conf. on Artificial Intelligence and Statistics
  ({AISTATS})}, pages 509--517, 2017.

\bibitem{velivckovic2017graph}
Petar Velickovic, Guillem Cucurull, Arantxa Casanova, Adriana Romero, Pietro
  Lio, and Yoshua Bengio.
\newblock Graph attention networks.
\newblock In {\em Int. Conf. on Learning Representations ({ICLR})}, 2018.

\bibitem{verbraeken2020survey}
Joost Verbraeken, Matthijs Wolting, Jonathan Katzy, Jeroen Kloppenburg, Tim
  Verbelen, and Jan~S Rellermeyer.
\newblock A survey on distributed machine learning.
\newblock {\em ACM Computing Surveys (CSUR)}, 53(2):1--33, 2020.

\bibitem{vishnu2016distributed}
Abhinav Vishnu, Charles Siegel, and Jeffrey Daily.
\newblock Distributed tensorflow with mpi.
\newblock {\em arXiv preprint arXiv:1603.02339}, 2016.

\bibitem{wainakh2020enhancing}
Aidmar Wainakh, Alejandro~Sanchez Guinea, Tim Grube, and Max
  M{\"u}hlh{\"a}user.
\newblock Enhancing privacy via hierarchical federated learning.
\newblock In {\em IEEE European Symposium on Security and Privacy Workshops
  (EuroS\&PW)}, pages 344--347, 2020.

\bibitem{wang2020graphfl}
Binghui Wang, Ang Li, Hai Li, and Yiran Chen.
\newblock Graphfl: A federated learning framework for semi-supervised node
  classification on graphs.
\newblock {\em arXiv preprint arXiv:2012.04187}, 2020.

\bibitem{Wang2021AGCNS}
Chunnan Wang, Bozhou Chen, Geng Li, and Hongzhi Wang.
\newblock {FL-AGCNS:} federated learning framework for automatic graph
  convolutional network search.
\newblock {\em arXiv preprint arXiv:2104.04141}, 2021.

\bibitem{wang2019interpret}
Guan Wang.
\newblock Interpret federated learning with shapley values.
\newblock {\em arXiv preprint arXiv:1905.04519}, 2019.

\bibitem{Wang2020Federated}
Hongyi Wang, Mikhail Yurochkin, Yuekai Sun, Dimitris Papailiopoulos, and
  Yasaman Khazaeni.
\newblock Federated learning with matched averaging.
\newblock In {\em Int. Conf. on Learning Representations ({ICLR})}, 2020.

\bibitem{Wang2021Guide}
Jianyu Wang, Zachary Charles, Zheng Xu, Gauri Joshi, H.~Brendan McMahan,
  Blaise~Ag{\"{u}}era y~Arcas, Maruan Al{-}Shedivat, Galen Andrew, Salman
  Avestimehr, Katharine Daly, Deepesh Data, Suhas~N. Diggavi, Hubert Eichner,
  Advait Gadhikar, Zachary Garrett, Antonious~M. Girgis, Filip Hanzely, Andrew
  Hard, Chaoyang He, Samuel Horvath, Zhouyuan Huo, Alex Ingerman, Martin Jaggi,
  Tara Javidi, Peter Kairouz, Satyen Kale, Sai~Praneeth Karimireddy, Jakub
  Kone{\v{c}}n{\'y}, Sanmi Koyejo, Tian Li, Luyang Liu, Mehryar Mohri, Hang Qi,
  Sashank~J. Reddi, Peter Richt{\'{a}}rik, Karan Singhal, Virginia Smith, Mahdi
  Soltanolkotabi, Weikang Song, Ananda~Theertha Suresh, Sebastian~U. Stich,
  Ameet Talwalkar, Hongyi Wang, Blake~E. Woodworth, Shanshan Wu, Felix~X. Yu,
  Honglin Yuan, Manzil Zaheer, Mi~Zhang, Tong Zhang, Chunxiang Zheng, Chen Zhu,
  and Wennan Zhu.
\newblock A field guide to federated optimization.
\newblock {\em arXiv preprint arXiv:2107.06917}, 2021.

\bibitem{wang2019matcha}
Jianyu Wang, Anit~Kumar Sahu, Zhouyi Yang, Gauri Joshi, and Soummya Kar.
\newblock Matcha: Speeding up decentralized {SGD} via matching decomposition
  sampling.
\newblock In {\em Indian Control Conference ({ICC})}, pages 299--300, 2019.

\bibitem{wang2021addressing}
Lixu Wang, Shichao Xu, Xiao Wang, and Qi~Zhu.
\newblock Addressing class imbalance in federated learning.
\newblock In {\em AAAI Conf. on Artificial Intelligence}, volume~35, pages
  10165--10173, 2021.

\bibitem{luping2019cmfl}
Luping WANG, Wei WANG, and LI~Bo.
\newblock {CMFL}: Mitigating communication overhead for federated learning.
\newblock In {\em IEEE Int. Conf. on Distributed Computing Systems (ICDCS)},
  pages 954--964, 2019.

\bibitem{wang2019beyond}
Zhibo Wang, Mengkai Song, Zhifei Zhang, Yang Song, Qian Wang, and Hairong Qi.
\newblock Beyond inferring class representatives: User-level privacy leakage
  from federated learning.
\newblock In {\em IEEE Conf. on Computer Communications ({INFOCOM})}, pages
  2512--2520, 2019.

\bibitem{FATE}
WeBank.
\newblock Federated ai technology enabler ({FATE}).
\newblock \url{https://github.com/FederatedAI/FATE}.
\newblock Online; accessed 16/02/2021.

\bibitem{FLWhitePaper}
WeBank.
\newblock Federated learning white paper v2.0.
\newblock
  \url{https://aisp-1251170195.cos.ap-hongkong.myqcloud.com/wp-content/uploads/pdf/\%E8\%81\%94\%E9\%82\%A6\%E5\%AD\%A6\%E4\%B9\%A0\%E7\%99\%BD\%E7\%9A\%AE\%E4\%B9\%A6_v2.0.pdf}.
\newblock Online; accessed 14/02/2021.

\bibitem{wei2020federated}
Kang Wei, Jun Li, Ming Ding, Chuan Ma, Howard~H Yang, Farhad Farokhi, Shi Jin,
  Tony~QS Quek, and H~Vincent Poor.
\newblock Federated learning with differential privacy: Algorithms and
  performance analysis.
\newblock {\em IEEE Transactions on Information Forensics and Security},
  15:3454--3469, 2020.

\bibitem{wu2021fedgnn}
Chuhan Wu, Fangzhao Wu, Yang Cao, Yongfeng Huang, and Xing Xie.
\newblock Fedgnn: Federated graph neural network for privacy-preserving
  recommendation.
\newblock {\em arXiv preprint arXiv:2102.04925}, 2021.

\bibitem{wu2021adversarial}
Tong Wu, Ziwei Liu, Qingqiu Huang, Yu~Wang, and Dahua Lin.
\newblock Adversarial robustness under long-tailed distribution.
\newblock In {\em IEEE/CVF Conference on Computer Vision and Pattern
  Recognition ({CVPR})}, pages 8659--8668, 2021.

\bibitem{xu2013survey}
Chang Xu, Dacheng Tao, and Chao Xu.
\newblock A survey on multi-view learning.
\newblock {\em arXiv preprint arXiv:1304.5634}, 2013.

\bibitem{xu2020federated}
Jie Xu, Benjamin~S Glicksberg, Chang Su, Peter Walker, Jiang Bian, and Fei
  Wang.
\newblock Federated learning for healthcare informatics.
\newblock {\em Journal of Healthcare Informatics Research}, pages 1--19, 2020.

\bibitem{xu2020ternary}
Jinjin Xu, Wenli Du, Yaochu Jin, Wangli He, and Ran Cheng.
\newblock Ternary compression for communication-efficient federated learning.
\newblock {\em IEEE Transactions on Neural Networks and Learning Systems},
  2020.

\bibitem{yang2019federated}
Qiang Yang, Yang Liu, Tianjian Chen, and Yongxin Tong.
\newblock Federated machine learning: Concept and applications.
\newblock {\em ACM Transactions on Intelligent Systems and Technology (TIST)},
  10(2):1--19, 2019.

\bibitem{yi2014homomorphic}
Xun Yi, Russell Paulet, and Elisa Bertino.
\newblock Homomorphic encryption.
\newblock In {\em Homomorphic Encryption and Applications}, pages 27--46.
  Springer, 2014.

\bibitem{yuan2020hierarchical}
Jinliang Yuan, Mengwei Xu, Xiao Ma, Ao~Zhou, Xuanzhe Liu, and Shangguang Wang.
\newblock Hierarchical federated learning through lan-wan orchestration.
\newblock {\em arXiv preprint arXiv:2010.11612}, 2020.

\bibitem{yurochkin2019bayesian}
Mikhail Yurochkin, Mayank Agarwal, Soumya Ghosh, Kristjan Greenewald, Nghia
  Hoang, and Yasaman Khazaeni.
\newblock Bayesian nonparametric federated learning of neural networks.
\newblock In {\em Int. Conf. on Machine Learning ({ICML})}, pages 7252--7261,
  2019.

\bibitem{zhang2020batchcrypt}
Chengliang Zhang, Suyi Li, Junzhe Xia, Wei Wang, Feng Yan, and Yang Liu.
\newblock Batchcrypt: Efficient homomorphic encryption for cross-silo federated
  learning.
\newblock In {\em {USENIX} Annual Technical Conference ({USENIX} {ATC})}, pages
  493--506, 2020.

\bibitem{zhang2017feature}
Chongsheng Zhang, Jingjun Bi, and Paolo Soda.
\newblock Feature selection and resampling in class imbalance learning: Which
  comes first? an empirical study in the biological domain.
\newblock In {\em Int. Conf. on Bioinformatics and Biomedicine ({BIBM})}, pages
  933--938, 2017.

\bibitem{zhang2019multi}
Chongsheng Zhang, Jingjun Bi, Shixin Xu, Enislay Ramentol, Gaojuan Fan, Baojun
  Qiao, and Hamido Fujita.
\newblock Multi-imbalance: An open-source software for multi-class imbalance
  learning.
\newblock {\em Knowledge-Based Systems}, 174:137--143, 2019.

\bibitem{zhang2021empirical}
Chongsheng Zhang, Paolo Soda, Jingjun Bi, Gaojuan Fan, George Almpanidis, and
  Salvador Garcia.
\newblock An empirical study on the joint impact of feature selection and data
  resampling on imbalance classification.
\newblock {\em arXiv preprint arXiv:2109.00201}, 2021.

\bibitem{Zhang2021FederatedGraph}
Huanding Zhang, Tao Shen, Fei Wu, Mingyang Yin, Hongxia Yang, and Chao Wu.
\newblock Federated graph learning - {A} position paper.
\newblock {\em arXiv preprint arXiv:2105.11099}, 2021.

\bibitem{zhang2021federated}
Tuo Zhang, Chaoyang He, Tianhao Ma, Mark Ma, and Salman Avestimehr.
\newblock Federated learning for internet of things: A federated learning
  framework for on-device anomaly data detection.
\newblock {\em arXiv preprint arXiv:2106.07976}, 2021.

\bibitem{zhang2020enabling}
Xiaoli Zhang, Fengting Li, Zeyu Zhang, Qi~Li, Cong Wang, and Jianping Wu.
\newblock Enabling execution assurance of federated learning at untrusted
  participants.
\newblock In {\em IEEE INFOCOM Conf. on Computer Communications}, pages
  1877--1886, 2020.

\bibitem{zhao2020idlg}
Bo~Zhao, Konda~Reddy Mopuri, and Hakan Bilen.
\newblock idlg: Improved deep leakage from gradients.
\newblock {\em arXiv preprint arXiv:2001.02610}, 2020.

\bibitem{zhao2021multimodal}
Yuchen Zhao, Payam Barnaghi, and Hamed Haddadi.
\newblock Multimodal federated learning.
\newblock {\em arXiv preprint arXiv:2109.04833}, 2021.

\bibitem{Zheng2021ASFGNN}
Longfei Zheng, Jun Zhou, Chaochao Chen, Bingzhe Wu, Li~Wang, and Benyu Zhang.
\newblock Asfgnn: Automated separated-federated graph neural network.
\newblock {\em Peer-to-Peer Networking and Applications}, 14(3):1692--1704,
  2021.

\bibitem{zhou2020vertically}
Jun Zhou, Chaochao Chen, Longfei Zheng, Huiwen Wu, Jia Wu, Xiaolin Zheng,
  Bingzhe Wu, Ziqi Liu, and Li~Wang.
\newblock Vertically federated graph neural network for privacy-preserving node
  classification.
\newblock {\em arXiv preprint arXiv:2005.11903}, 2020.

\bibitem{zhu2021}
Hangyu Zhu, Haoyu Zhang, and Yaochu Jin.
\newblock From federated learning to federated neural architecture search: a
  survey.
\newblock {\em Complex \& Intelligent Systems}, 2021.

\bibitem{zinkevich2010parallelized}
Martin Zinkevich, Markus Weimer, Alexander~J Smola, and Lihong Li.
\newblock Parallelized stochastic gradient descent.
\newblock In {\em Advances in Neural Information Processing Systems
  ({NeurIPS})}, volume~4, page~4. Citeseer, 2010.

\end{thebibliography}

\end{document}